\documentclass[twoside,twocolumn,9pt]{article}
\usepackage{extsizes}
\usepackage[super,sort&compress,comma]{natbib} 
\usepackage[version=3]{mhchem}
\usepackage[left=1.5cm, right=1.5cm, top=1.785cm, bottom=2.0cm]{geometry}
\usepackage{balance}
\usepackage{mathptmx}
\usepackage{sectsty}
\usepackage{graphicx} 
\usepackage{lastpage}
\usepackage[format=plain,justification=justified,singlelinecheck=false,font={stretch=1.125,small,sf},labelfont=bf,labelsep=space]{caption}
\usepackage{float}
\usepackage{fancyhdr}
\usepackage{fnpos}
\usepackage[english]{babel}
\addto{\captionsenglish}{%
  
}
\usepackage{array}
\usepackage{droidsans}
\usepackage{charter}
\usepackage[T1]{fontenc}
\usepackage[usenames,dvipsnames]{xcolor}
\usepackage{setspace}
\usepackage[compact]{titlesec}
\usepackage{hyperref}

\usepackage{subcaption}
\usepackage{amssymb}
\usepackage{xcolor}
\usepackage{comment}
\excludecomment{toexclude}

\usepackage{epstopdf}

\definecolor{cream}{RGB}{222,217,201}

\begin{document}

\pagestyle{fancy}
\thispagestyle{plain}
\fancypagestyle{plain}{
\renewcommand{\headrulewidth}{0pt}
}

\makeFNbottom
\makeatletter
\renewcommand\LARGE{\@setfontsize\LARGE{15pt}{17}}
\renewcommand\Large{\@setfontsize\Large{12pt}{14}}
\renewcommand\large{\@setfontsize\large{10pt}{12}}
\renewcommand\footnotesize{\@setfontsize\footnotesize{7pt}{10}}
\makeatother

\renewcommand{\thefootnote}{\fnsymbol{footnote}}
\renewcommand\footnoterule{\vspace*{1pt}%
\color{cream}\hrule width 3.5in height 0.4pt \color{black}\vspace*{5pt}} 
\setcounter{secnumdepth}{5}

\makeatletter 
\renewcommand\@biblabel[1]{#1}            
\renewcommand\@makefntext[1]%
{\noindent\makebox[0pt][r]{\@thefnmark\,}#1}
\makeatother 
\renewcommand{\figurename}{\small{Fig.}~}
\sectionfont{\sffamily\Large}
\subsectionfont{\normalsize}
\subsubsectionfont{\bf}
\setstretch{1.125} 
\setlength{\skip\footins}{0.8cm}
\setlength{\footnotesep}{0.25cm}
\setlength{\jot}{10pt}
\titlespacing*{\section}{0pt}{4pt}{4pt}
\titlespacing*{\subsection}{0pt}{15pt}{1pt}

\fancyfoot{}
\fancyfoot[LO,RE]{\vspace{-7.1pt}\includegraphics[height=9pt]{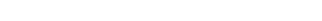}}
\fancyfoot[CO]{\vspace{-7.1pt}\hspace{11.9cm}\includegraphics{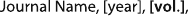}}
\fancyfoot[CE]{\vspace{-7.2pt}\hspace{-13.2cm}\includegraphics{head_foot/RF}}
\fancyfoot[RO]{\footnotesize{\sffamily{1--\pageref{LastPage} ~\textbar  \hspace{2pt}\thepage}}}
\fancyfoot[LE]{\footnotesize{\sffamily{\thepage~\textbar\hspace{4.65cm} 1--\pageref{LastPage}}}}
\fancyhead{}
\renewcommand{\headrulewidth}{0pt} 
\renewcommand{\footrulewidth}{0pt}
\setlength{\arrayrulewidth}{1pt}
\setlength{\columnsep}{6.5mm}
\setlength\bibsep{1pt}

\makeatletter 
\newlength{\figrulesep} 
\setlength{\figrulesep}{0.5\textfloatsep} 

\newcommand{\topfigrule}{\vspace*{-1pt}%
\noindent{\color{cream}\rule[-\figrulesep]{\columnwidth}{1.5pt}} }

\newcommand{\botfigrule}{\vspace*{-2pt}%
\noindent{\color{cream}\rule[\figrulesep]{\columnwidth}{1.5pt}} }

\newcommand{\dblfigrule}{\vspace*{-1pt}%
\noindent{\color{cream}\rule[-\figrulesep]{\textwidth}{1.5pt}} }

\makeatother

\twocolumn[
  \begin{@twocolumnfalse}
{\includegraphics[height=30pt]{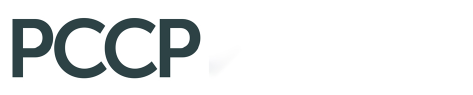}\hfill\raisebox{0pt}[0pt][0pt]{\includegraphics[height=55pt]{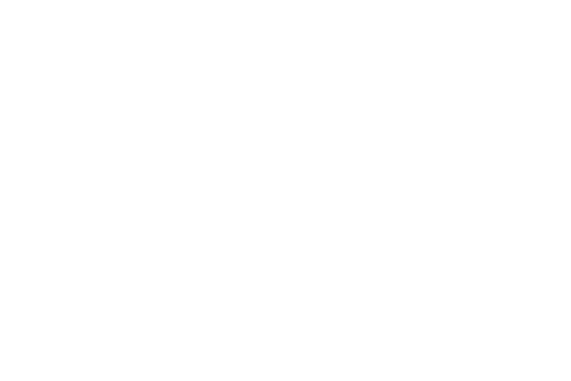}}\\[1ex]
\includegraphics[width=18.5cm]{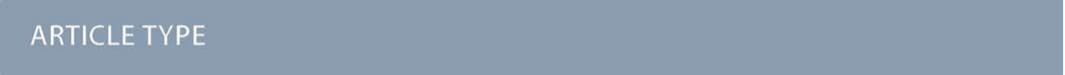}}\par
\vspace{1em}
\sffamily
\begin{tabular}{m{4.5cm} p{13.5cm} }

\includegraphics{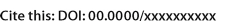} & \noindent\LARGE{\textbf{Surface phase diagrams from nested sampling$^\dag$}} \\
\vspace{0.3cm} & \vspace{0.3cm} \\

 & \noindent\large{Mingrui Yang,\textit{$^{a}$} Livia B. P\'{a}rtay,\textit{$^{b}$} and Robert B. Wexler$^{\ast}$\textit{$^{a}$}} \\

\includegraphics{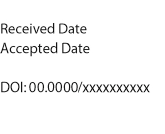} & \noindent\normalsize{Studies in atomic-scale modeling of surface phase equilibria often focus on temperatures near zero Kelvin due to the challenges in calculating the free energy of surfaces at finite temperatures. The Bayesian-inference-based nested sampling (NS) algorithm allows for modeling phase equilibria at arbitrary temperatures by directly and efficiently calculating the partition function, whose relationship with free energy is well known. This work extends NS to calculate adsorbate phase diagrams, incorporating all relevant configurational contributions to the free energy. We apply NS to the adsorption of Lennard-Jones (LJ) gas particles on low-index and vicinal LJ solid surfaces and construct the canonical partition function from these recorded energies to calculate ensemble averages of thermodynamic properties, such as the constant-volume heat capacity and order parameters that characterize the structure of adsorbate phases. Key results include determining the nature of phase transitions of adsorbed LJ particles on flat and stepped LJ surfaces, which typically feature an enthalpy-driven condensation at higher temperatures and an entropy-driven reordering process at lower temperatures, and the effect of surface geometry on the presence of triple points in the phase diagrams. Overall, we demonstrate the ability and potential of NS for surface modeling.} \\

\end{tabular}

 \end{@twocolumnfalse} \vspace{0.6cm}

  ]

\renewcommand*\rmdefault{bch}\normalfont\upshape
\rmfamily
\section*{}
\vspace{-1cm}


\footnotetext{\textit{$^{a}$~Department of Chemistry and Institute of Materials Science and Engineering, Washington University in St. Louis, St. Louis, MO 63130, USA. E-mail: wexler@wustl.edu}}
\footnotetext{\textit{$^{b}$~Department of Chemistry, University of Warwick, Coventry, CV4 7AL, UK.}}
\footnotetext{\dag~Electronic Supplementary Information (ESI) available: finite-size analysis, system setup procedures, specifications of surface structure, calculations of order parameters, and illustrations of maximum-probability surface structures. See DOI: \url{https://doi.org/10.1039/cXCP00000x}}


\section{Introduction}

The structure and composition of solid surfaces play an essential role in determining their properties for various high-stakes applications, including gas sensing \cite{wander_stability_2001} and optoelectronics, \cite{ravi_origin_2017} in addition to those central to mitigating the effects of climate change, such as photovoltaics \cite{ohmann_real-space_2015} and catalysis. For the latter, studies show that \textit{operando} surface reconstruction, \textit{i.e.}, changes in the structure and composition of the topmost layers of a solid, can govern the activity and selectivity of heterogeneous catalysts for the \ce{CO2} reduction, \cite{chen_self-assembling_2019, phan_emergence_2021} \ce{H2} evolution, \cite{cheng_restructuring_2023, wan_amorphous_2023, zhang_hydrogen-induced_2022} \ce{O2} evolution, \cite{martirez_synergistic_2015, weber_atomistic_2022, akbashev_probing_2023} and \ce{O2} reduction reactions, \cite{schroder_tracking_2023} to name a few examples. Therefore, designing catalysts for these reactions and functional materials for other surface-specific processes requires an atomic-scale understanding of surface reconstruction and its dependence on operating conditions. However, the experiments that can measure these properties are typically done under conditions that differ from operating conditions, \cite{akbashev_electrocatalysis_2022} with only a few exceptions. \cite{grumelli_electrochemical_2020}

To fulfill the need for atomic-scale information about \textit{operando} surface phases, the field has often turned to computer simulations, which typically fall under one of three categories: thermodynamic, optimization-based, and statistical thermodynamic approaches. The most notable example of the former is \textit{ab initio} surface thermodynamics, pioneered by Scheffler and coworkers. \cite{wang_the_1998, wang_effect_2000, reuter_composition_2001} In this approach, one uses quantum-mechanics-based simulation techniques to calculate the grand potentials per unit surface area for a set of slabs, the structures and compositions of which are typically guided by chemical and physical intuition (see Figure~\ref{fig:surface-setup} for an example of a surface slab model). The resulting surface grand potentials are usually approximate because only those finite-temperature effects associated with the enthalpy and entropy of harmonic vibrations are included, if at all. The contributions due to anharmonic vibrations and, more generally, configurational degrees of freedom are commonly ignored \cite{zhou_determining_2019} (except for gas phases, where reference measurements or the ideal gas equation of state have been used) to make surface calculations computationally tractable, especially when employing an \textit{ab-initio}-based description of the chemical bonds, such as density functional theory.

While \textit{ab initio} thermodynamics can be insightful, in the absence of input from careful measurements, the variety of possible surface structures one must intuit is unclear. Optimization-based approaches have shown promise in more comprehensively navigating the ambiguity of surface phase space. These include enumerating mean-field configurations, \cite{ulissi_automated_2016} random structure searching, \cite{schusteritsch_predicting_2014} simulated annealing, \cite{timmermann_iro_2020} basin hopping, \cite{ronne_atomistic_2022} global activity search, \cite{sun_global_2021} stochastic surface walking, \cite{chen_automated_2022} evolutionary/genetic algorithms, \cite{zhu_evolutionary_2013, chiriki_constructing_2019, wang_finite-temperature_2020, wanzenbock_neural-network-backed_2022} particle swarm optimization, \cite{lu_self-assembled_2014} active learning, \cite{bisbo_efficient_2020} and reinforcement learning. \cite{jorgensen_atomistic_2019} The methods above guide the search for structures that minimize the zero-Kelvin surface energy as a function of composition, which are used as inputs for \textit{ab initio} thermodynamics in the grand canonical ensemble. Recent studies have increasingly employed (Markov chain) Monte Carlo (MC) simulations in various ensembles, such as canonical, \cite{zhang_ensembles_2020} grand canonical, \cite{fantauzzi_growth_2017, wexler_automatic_2019, somjit_atomic_2022, xu_atomistic_2022} and semi-grand canonical, \cite{du_machine-learning-accelerated_2023} to facilitate the discovery of \textit{operando} surface phases. These simulations leverage the Metropolis-Hastings algorithm, allowing efficient sampling from specific probability distributions within defined thermodynamic constraints. This approach is particularly effective for potential energy surfaces (PESs) characterized by shallow wells. However, it encounters limitations in systems with deep potential wells, where ergodicity can be broken. This results in simulations that heavily depend on initial conditions and may not comprehensively explore configuration space. Consequently, accurately calculating the free energy in such systems remains a complex task. In these scenarios, generating accurate surface phase diagrams is challenging unless it is established that entropic contributions are minimal.

Recently, Zhou, Scheffler, and Ghiringhelli introduced an approach to efficiently estimate the free energy of surfaces from replica-exchange (RE) grand-canonical sampling and the multi-state Bennett acceptance ratio method. \cite{zhou_determining_2019} Their algorithm marks a notable advancement in computational surface science, especially with the inclusion of anharmonic contributions to free energy. It is interesting to consider that calculating free energy involves an integral over the phase space volume. Therefore, employing a sampling grid based on phase space volume, in contrast to the temperature-based approach used in RE and the energy-based approach used in Wang-Landau sampling, \cite{wang_efficient_2001, wang_determining_2001} could provide a more accurate and efficient approach to sampling near and away from surface phase transitions. This method offers the advantage of not requiring prior knowledge of the stable phases, \cite{partay_efficient_2010} further enhancing the already significant capabilities of the RE approach. We propose an alternative approach based on nested sampling (NS), which constructs a set of slabs equidistant in the natural logarithm of surface phase space volume. NS was first introduced in Bayesian statistics, \cite{skilling_nested_2004, skilling_nested_2006} and was later adopted by various research fields \cite{ashton_nested_2022} and adapted to sample the PES of atomic-scale systems. \cite{partay_efficient_2010, baldock_determining_2016, partay_nested_2021} The power of NS has been demonstrated in studying various systems, including the formation of clusters, \cite{rossi_thermodynamics_2018, dorrell_thermodynamics_2019} calculation of the quantum partition function, \cite{szekeres_direct_2018} sampling transition paths, \cite{bolhuis_nested_2018} as well as computing the pressure-temperature phase diagram for various metals, alloys, and model potentials, \cite{baldock_constant-pressure_2017, gola_embedded_2018, dorrell_pressuretemperature_2020, bartok_insight_2021, marchant2023exploring} which often identified previously unknown stable solid phases.

This study uses NS to calculate coverage-temperature adsorbate phase diagrams, a subset of surface phase diagrams, as a proof of concept for future investigations of more complex material interfaces and interfacial conditions. Here, ``coverage'' follows the standard definition as the ratio between the number of adsorbed particles on a surface and the number of particles in a filled monolayer (ML) on that surface. We carry out NS for surfaces of the Lennard-Jones (LJ) solid, constructing the canonical partition function from potential energy values recorded while compressing the natural logarithm of the adsorbate phase space volume by a constant factor at each iteration. While a considerable body of research exists on the LJ system, much of this work has concentrated on its bulk solid (for example, see References \citenum{sprik_second_1984, van_de_waal_can_1991, van_der_hoef_free_2000}) and fluid phases (for example, see References \citenum{nicolas_equation_1979, smit_phase_1992, mecke_molecular_1997}), as well as the physical properties of its bulk-terminated surfaces (for example, see References \citenum{broughton_surface_1983, valkealahti_molecular_1987, somasi_computer_2001}). Significant progress has been made in understanding these aspects.

Nonetheless, the comprehensive examination of coverage-temperature adsorbate phase diagrams for LJ solids, especially beyond two-dimensional models and simpler lattice frameworks, presents an opportunity for further exploration. This aspect has not been as extensively covered as other areas, as seen in the more focused research that primarily addressed two-dimensional cases (for example, see References \citenum{berker_renormalization-group_1978, ostlund_structural_1980, patrykiejew_monte_1986, borowski_two-dimensional_1989, patrykiejew_two-dimensional_1991}). Our approach differs by sampling a continuum of particle positions within the canonical ensemble rather than discretized particle positions (\textit{i.e.}, a lattice gas model) in the grand canonical ensemble. \cite{lee_statistical_1952, doyen_orderdisorder_1975, binder_multicritical_1976} While the grand canonical ensemble offers insights more directly pertinent to catalyst design, employing the canonical ensemble is more computationally efficient. This efficiency is particularly beneficial in reducing the computational demands associated with benchmarking surface NS. Our work thus contributes to a more nuanced understanding of the LJ system, complementing existing research while exploring new and potentially fruitful areas.

Utilizing the constructed canonical partition function, we compute ensemble averages of thermodynamic properties such as the constant-volume heat capacity and order parameters that describe the structure of adsorbate phases. For simplicity, the particles in the solid are not allowed to move; thus, the surface is not allowed to relax or reconstruct under the influence of the adsorbed particles. Such simplification ensures that only the direct interplay between adsorbed particles and the surface is captured, providing a clearer picture of the adsorption processes. More details are included in Section~\ref{computational-methods}. Notably, we identify phase transitions of the adsorbed particles on both flat and stepped surfaces. Most flat surfaces exhibit an enthalpy-driven condensation at higher temperatures, an entropy-driven reordering process within the condensed layer at lower temperatures, and a coverage above which a disordered adsorbate phase is unstable. Ultimately, we showcase the capabilities and potential of NS for surface modeling.

\section{Computational methods} \label{computational-methods}

\subsection{Nested sampling}

\subsubsection{Algorithm}

NS is an algorithm for computing Bayesian evidence, which, in statistical mechanics, can be analogously related to the partition function. Independent configurations of particles, commonly referred to as ``walkers'' or ``live points,'' are employed to sample atomic configuration space. \cite{partay_efficient_2010} The number of walkers, $K$, remains constant during the sampling and determines the sampling resolution. NS is started by drawing a set of random walkers from a uniform prior distribution in configuration space, \textit{i.e.}, a set of ideal-gas-like configurations. These configurations establish a range of potential energies, the negative exponentials of which are analogous to their Bayesian likelihoods. During each iteration of NS, the walker with the highest potential energy (\textit{i.e.}, the least likelihood value) is identified, the contribution of the sampled configuration to the canonical partition function (\textit{i.e.}, the evidence) is calculated, and this configuration and its potential energy are recorded. Then, this walker is replaced by a new walker drawn from the set of ideal-gas-like configurations (\textit{i.e.}, the prior distribution) but with a constraint: its potential energy must be less than that of the replaced configuration. This process is repeated until the lowest potential energy configuration is identified. The idea is that the distribution of configurations is ``narrowed down'' or ``nested'' into regions of decreasing potential energy. This approach provides an efficient way of exploring configuration spaces where most low-potential-energy configurations are located within a small fraction of the phase space volume, a common occurrence in systems that follow the Boltzmann distribution.

Previous studies have detailed NS, outlining its use in studying clusters at constant volume \cite{partay_efficient_2010, martiniani_superposition_2014} and bulk materials at constant pressure \cite{baldock_determining_2016, partay_nested_2021}; here, we will concentrate on adsorbates at constant volume. In our study, we differentiate between particles that make up the surface and those that adsorb or condense onto it. The adsorbing particles, which we term \textit{free} particles, can move and change position, even when condensed on the surface. In contrast, \textit{fixed} particles refer to those in the fixed structure of the surface, which do not change position during sampling. This distinction helps to simplify our model system, focusing on the key interactions between mobile free particles and the static surface. With the primary objective of establishing proof of concept, we have prioritized simplicity to eliminate potential complexities that could obscure the phenomena of interest.

Once the initial walkers are generated, the iterative part of the sampling proceeds as follows:

\begin{enumerate}
    \item \label{step1} Identify and record the walker with the highest potential energy, denoted as $E_i^{\max} = \max{\{ E \}}$. The phase space volume, $\Gamma$, of the PES below $E_i^{\max}$ is given by $\Gamma_i = \left[ K / \left( K + 1 \right) \right]^i$. \cite{partay_efficient_2010}
    \item \label{step2} Replace the highest-potential-energy walker with a new walker, where the particle positions of the new configuration are chosen randomly, but such that $E^{\mathrm{new}} < E_i^{\max}$. As configurations with lower potential energy are sampled, the available phase space volume shrinks. Consequently, the probability of generating an acceptable random walker diminishes. Hence, we generate a new walker by cloning a randomly selected existing walker. We then employ rejection sampling to conduct a random walk in configuration space, ensuring the walker's potential energy remains below $E_i^{\max}$. This process effectively decorrelates the new configuration, making it statistically independent of its starting point.
    \item \label{step3} Let $i \leftarrow i + 1$ and return to Step~\ref{step1}.
\end{enumerate}

Once sufficiently low-potential-energy regions of configuration space are explored, NS can stop, and using the set of recorded potential energy values from the replaced walkers, we can calculate the canonical partition function, $Z$, as

\begin{equation} \label{eq: Z}
    Z \left( N, V, \beta \right) = \sum_i w_i \exp{\left[ - \beta E_i \left( N, V \right) \right]},
\end{equation}

\noindent where $N$ is the number of particles, $V$ is the volume, $\beta = 1 / k_{\mathrm{B}} T $ is the inverse temperature, and $w_i$ is the NS weight of the $i$-th iteration, $w_i = \Gamma_i - \Gamma_{i + 1}$. \cite{partay_efficient_2010} Since NS (see Steps~\ref{step1}-\ref{step3} above) is temperature-independent (temperature is only considered in post-processing steps), one can substitute any $\beta$ into Equation~\ref{eq: Z} and therefore calculate the partition function at any temperature. Owing to its ``top-down'' approach, NS obviates the need for pre-existing knowledge about structural or thermodynamic properties. This feature renders it particularly effective for an unbiased and comprehensive exploration of the PES, making it highly suitable for identifying thermodynamically relevant phases. We focus on phase transitions, identifiable through local extrema and inflection points in thermodynamic properties derived from $Z$. These include peaks in the heat capacity that are not influenced by the scale of potential energy. In our calculations, we focus solely on the kinetic contributions of the free particles, treating them classically in post-processing steps while keeping the surface particles fixed.

\subsubsection{Phase equilibria\label{sec:phase-equilibria}}

The presence of a peak in the heat capacity, $C_V$, is suggestive of a phase transition and hence we use this to locate transitions. However, in the current work we do not speculate on whether the observed transitions are first-order- or continuous-like in the studied finite systems.
We calculate $C_V$ as a function of $\beta$, whose relationship with $Z$ is well known, \textit{i.e.},

\begin{equation}
    C_V \left( \beta \right) = k_{\mathrm{B}} \beta^2 \left( \frac{\partial^2 \ln{Z}}{\partial \beta^2} \right)_{N, V}.
\end{equation}

\noindent To determine the peak positions on the $C_V \left( \beta \right)$ curve -- and hence the adsorbate phase transition temperatures -- for different coverages, we used the \verb|scipy.signal.find_peaks()| function with \verb|prominence=0.02| to automatically find peaks for each $C_V$ curve. We then manually connected adjacent peaks for neighboring coverages to construct a coverage-temperature phase diagram.

With access to the canonical partition function, we can also calculate the ensemble average of any configuration-dependent property, $A \left( \mathbf{r}_i \right)$, at a given temperature, $\beta$, as

\begin{equation} \label{eq: obs}
    \langle A \left( \beta \right) \rangle = \frac{\sum_i A \left( \mathbf{r}_i \right) w_i \exp{\left( - \beta E_i \right)}}{Z \left( \beta \right)}.
\end{equation}

\noindent We calculate surface order parameters to gain insight into the structural phase transitions of LJ surfaces. These include the average vertical position of the free particles relative to that of the fixed surface particles, $\langle \Delta z \rangle = \langle z_{\mathrm{free}} \rangle - z_{\mathrm{surface}}$, and the average coordination number of the free particles, $\langle \mathrm{CN} \rangle$, including free-free and free-fixed particle-particle bonds.

The probability $P_i$ of sampling configuration $i$ from the canonical ensemble is used to identify representative equilibrium structures of the system at a specific temperature. The calculation is performed using the following equation:

\begin{equation}
    P_i \left( \mathbf{r}_i, \beta \right) = \frac{w_i \exp{\left[ -\beta E \left( \mathbf{r}_i \right) \right]}}{Z \left( \beta \right)}.
\end{equation}

\noindent In this equation, $E \left( \mathbf{r}_i \right)$ is used to emphasize that the potential energy $E$ is determined by the particle positions $\mathbf{r}$ of configuration $i$. The structure that maximizes $P_i$ at a given temperature $\beta$ is thus the most likely to occur at that temperature. However, it is important to note that there can be numerous structures with probabilities close to the maximum, especially in stable high-entropy phases and at phase boundaries. Leveraging the recorded potential energy values for the replaced walkers in the NS method could also allow for constructing an optimal ensemble representation of fluxional catalytic interfaces \cite{zhang_ensembles_2020, poths_theoretical_2022, lavroff_ensemble_2022} in the thermodynamic limit.

\subsection{Simulation details \label{sec:sim-details}}

\subsubsection{Lennard-Jones potential}

In this study, we use the well-explored LJ potential \cite{stephan_thermophysical_2019} to demonstrate the capability of NS for predicting adsorbate phase diagrams. The LJ potential provides a simple yet physically meaningful model for surfaces interacting through spherically symmetric van der Waals-type forces. The LJ potential is typically expressed as

\begin{equation}
    V_{\mathrm{LJ}} \left( r \right) = 4 \epsilon \left[ \left( \frac{\sigma}{r} \right)^{12} - \left( \frac{\sigma}{r} \right)^{6} \right],
\end{equation}

\noindent where $\epsilon$ and $\sigma$ serve as the potential energy and distance units, respectively. We use the same $\sigma$ and $\epsilon$ values for free-free, free-fixed, and fixed-fixed particle interactions to model a pure LJ solid. We shifted the LJ potential to ensure the potential energy was zero at the cutoff radius of $4\sigma$.

\subsubsection{Surface system setup \label{sec:surface-setup}}

\begin{figure}[t]
    \centering
    \includegraphics[width=0.4\textwidth]{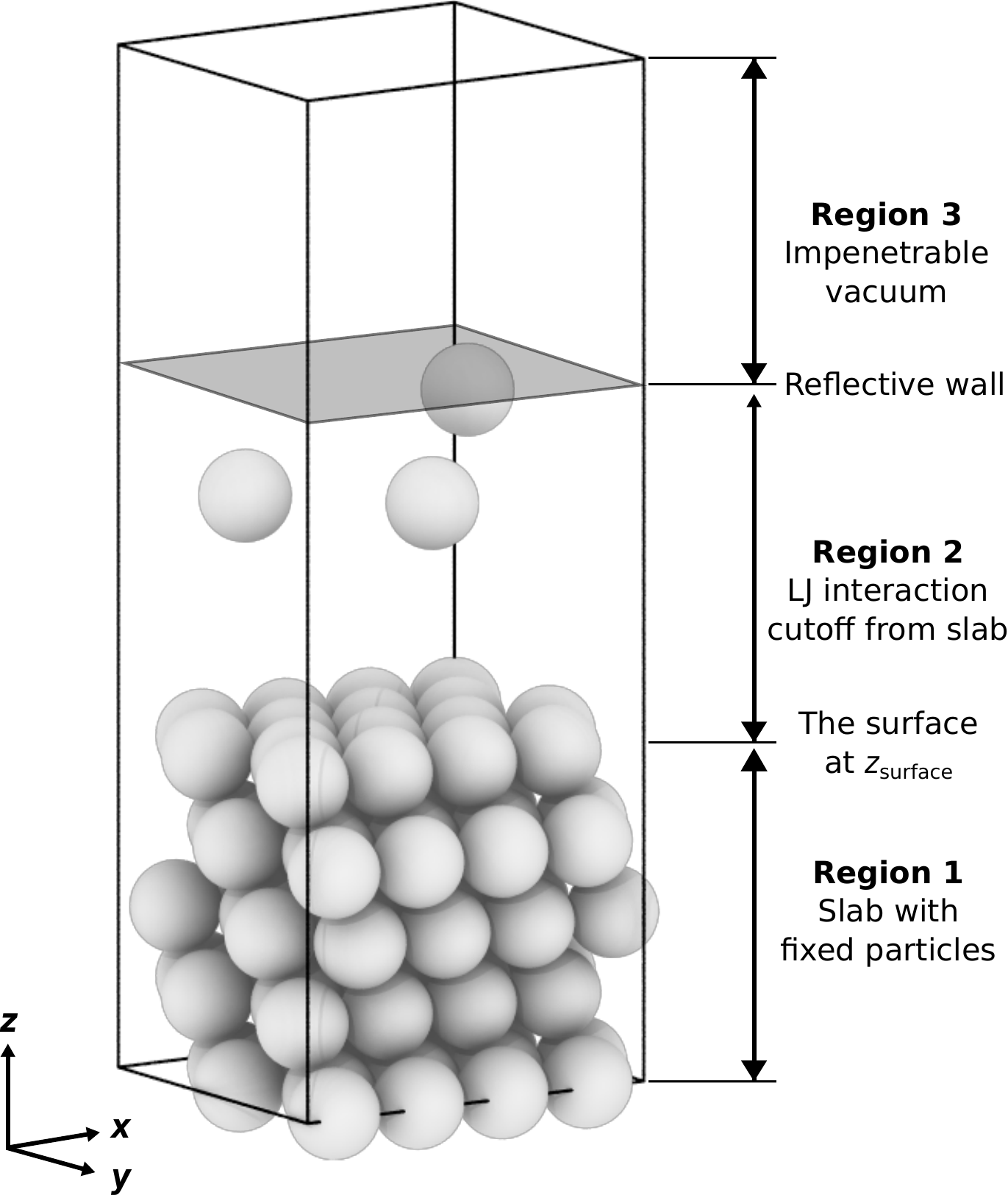}
    \caption{
An example of the surface system setup, showing a five-layer LJ(111) slab with a $4\times4$ surface unit cell (16 fixed particles per layer) and three free particles, corresponding to a maximum possible coverage of $\theta=3/16$ ML. From bottom to top, the three regions are as follows: (1) a slab with fixed particles and a thickness of approximately $4\sigma$, where the topmost monolayer defines the vertical position of the surface, $z_{\mathrm{surface}}$; (2) a sampled region where free particles can interact with the slab and other free particles; and (3) an approximately $4\sigma$-thick impenetrable vacuum to prevent the free particles from interacting with the periodic image of the bottom of the slab.
    }
    \label{fig:surface-setup}
\end{figure}

To sample the phase space of free LJ particles above a fixed LJ surface, we divide a simulation cell with three periodic dimensions into three regions from bottom to top (see Figure~\ref{fig:surface-setup} for an example of the setup):

\begin{itemize}
    \item \textbf{Region 1}: A slab with fixed particles and a thickness of approximately $4\sigma$, depending on the surface features, \textit{e.g.}, flat or stepped. The slab comprises several layers, each containing an identical count of particles. These layers are characterized by a fixed interlayer spacing, which varies based on the surface features. For a more detailed and quantitative description of different slab geometries, refer to Table~\ref{tab: SI_facets_params} in the ESI.
    \item \textbf{Region 2}: A space that extends $4 \sigma$ above the fixed slab, where the free particles can interact with the fixed slab and each other. The initial walkers are randomly and uniformly drawn from configurations in this region. We limit the space to a thickness of $\leq 4 \sigma$ (\textit{i.e.}, the value of the LJ cutoff radius) to exclude any space where the free particles do not interact with the slab (we call these ``voids,'' see Section~\ref{sec:surface-setup-details}). For systems having only one or two free particles or a disproportionately large gas phase region where the particles fall outside the interaction range of the surface, the sampling can struggle to find configurations with lower energies. In this case, additional measures could be necessary to ensure that NS proceeds to subsequent iterations by avoiding regions of configuration space where neighboring configurations have the same potential energy due to being outside the range of interaction with the surface. We describe such scenarios and solutions in Section~\ref{sec:surface-setup-details} of the ESI.

    \item \textbf{Region 3}: The topmost region is an impenetrable vacuum created by a reflective boundary located $4 \sigma$ below the cell's top. This boundary prevents the free particles from interacting with the slab's bottom due to periodic boundary conditions. Alternative boundary conditions, such as fluctuating boundaries, will be explored in future studies. These aim to eliminate the constraints imposed by the simulation cell's size, thus preventing artificial commensurability. They achieve this by introducing a phase shift at the boundary (also an MC simulation variable). \cite{olsson_a_1994}
\end{itemize}

\begin{figure*}[htb]
    \centering
    \begin{subfigure}[t]{0.24\textwidth}
        \caption{LJ(111) surface top view}
        \includegraphics[width=\textwidth]{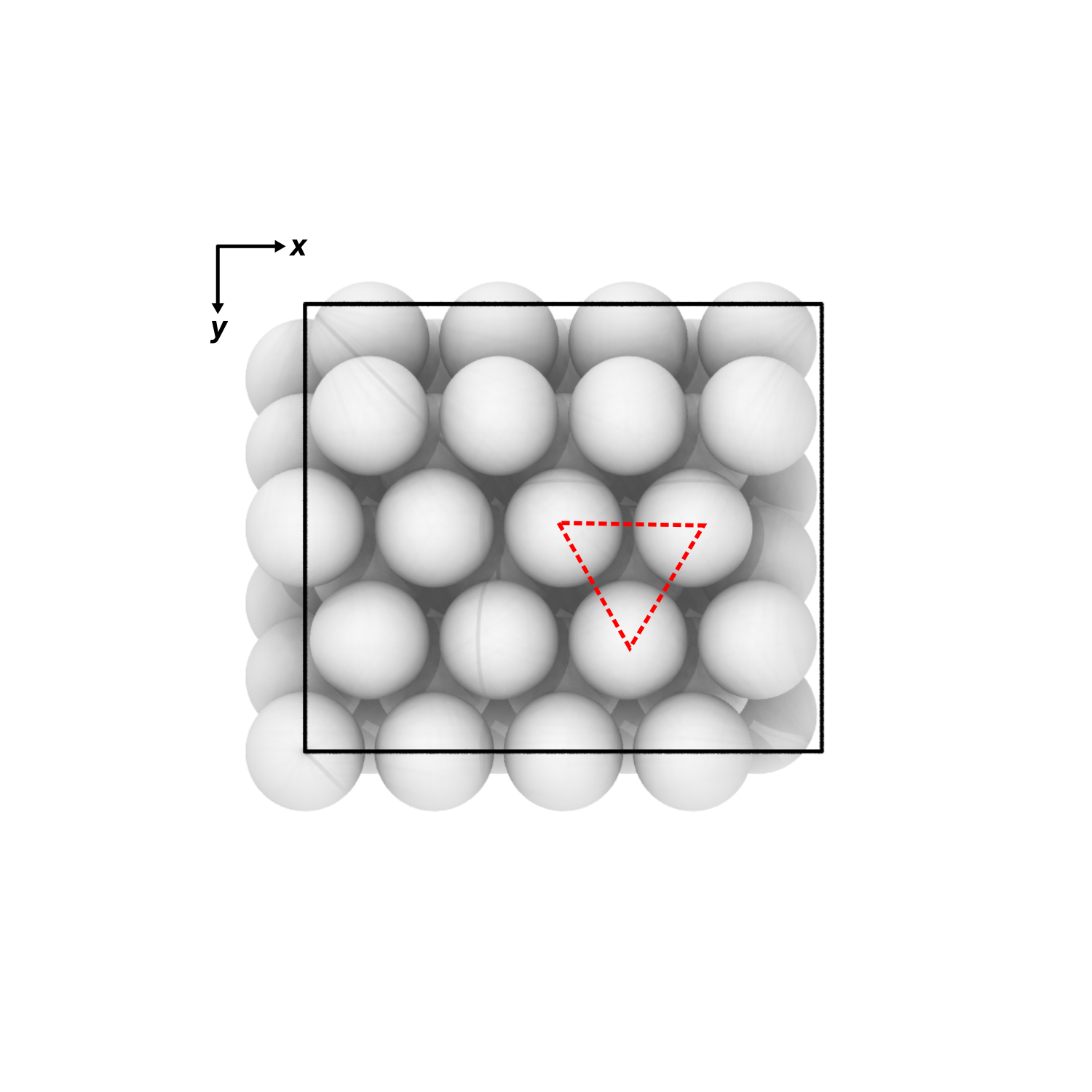}
        \label{fig:top111}
    \end{subfigure}
    \begin{subfigure}[t]{0.24\textwidth}
        \caption{LJ(100) surface top view}
        \includegraphics[width=\textwidth]{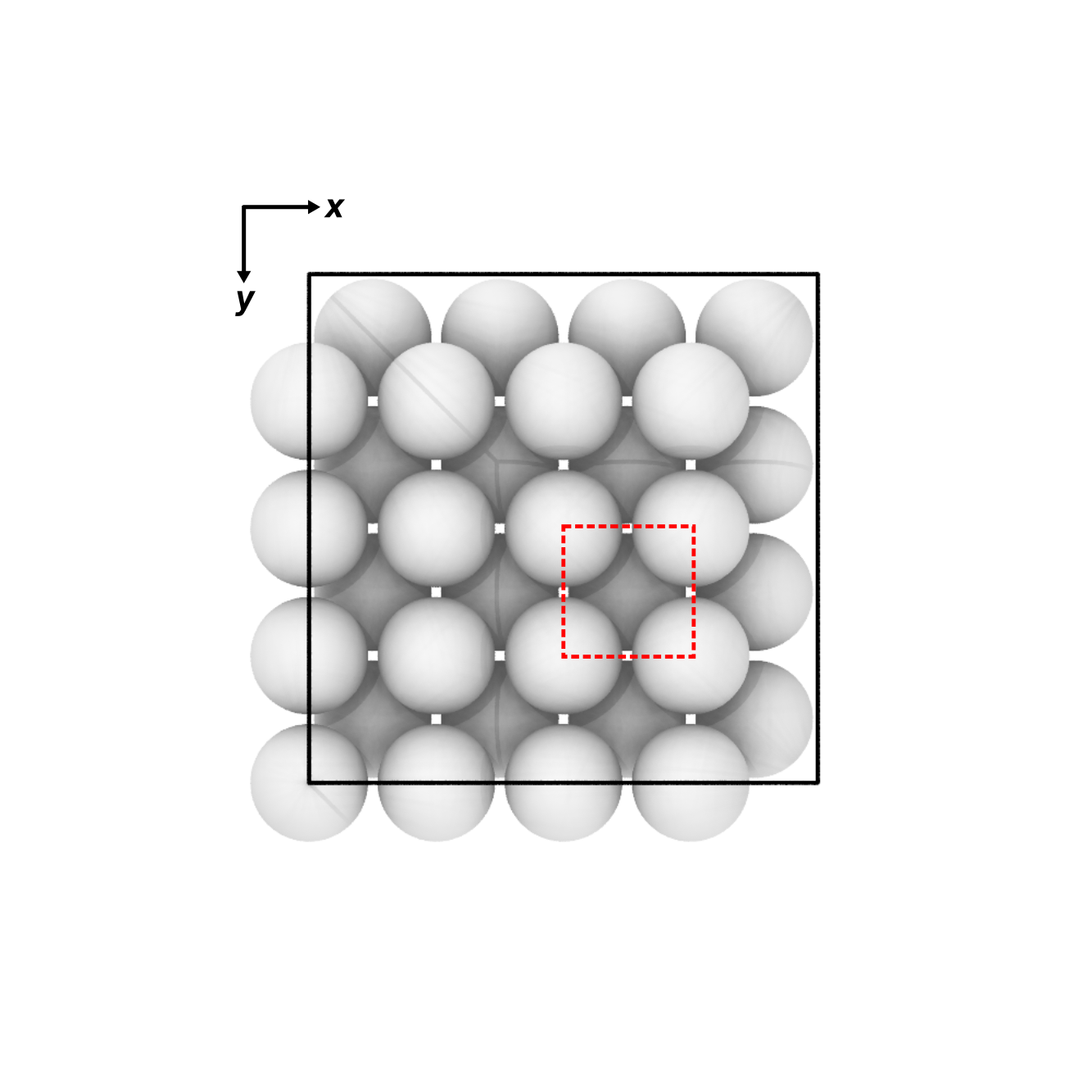}
        \label{fig:top100}
    \end{subfigure}
    \begin{subfigure}[t]{0.24\textwidth}
        \caption{LJ(311) surface top view}
        \includegraphics[width=\textwidth]{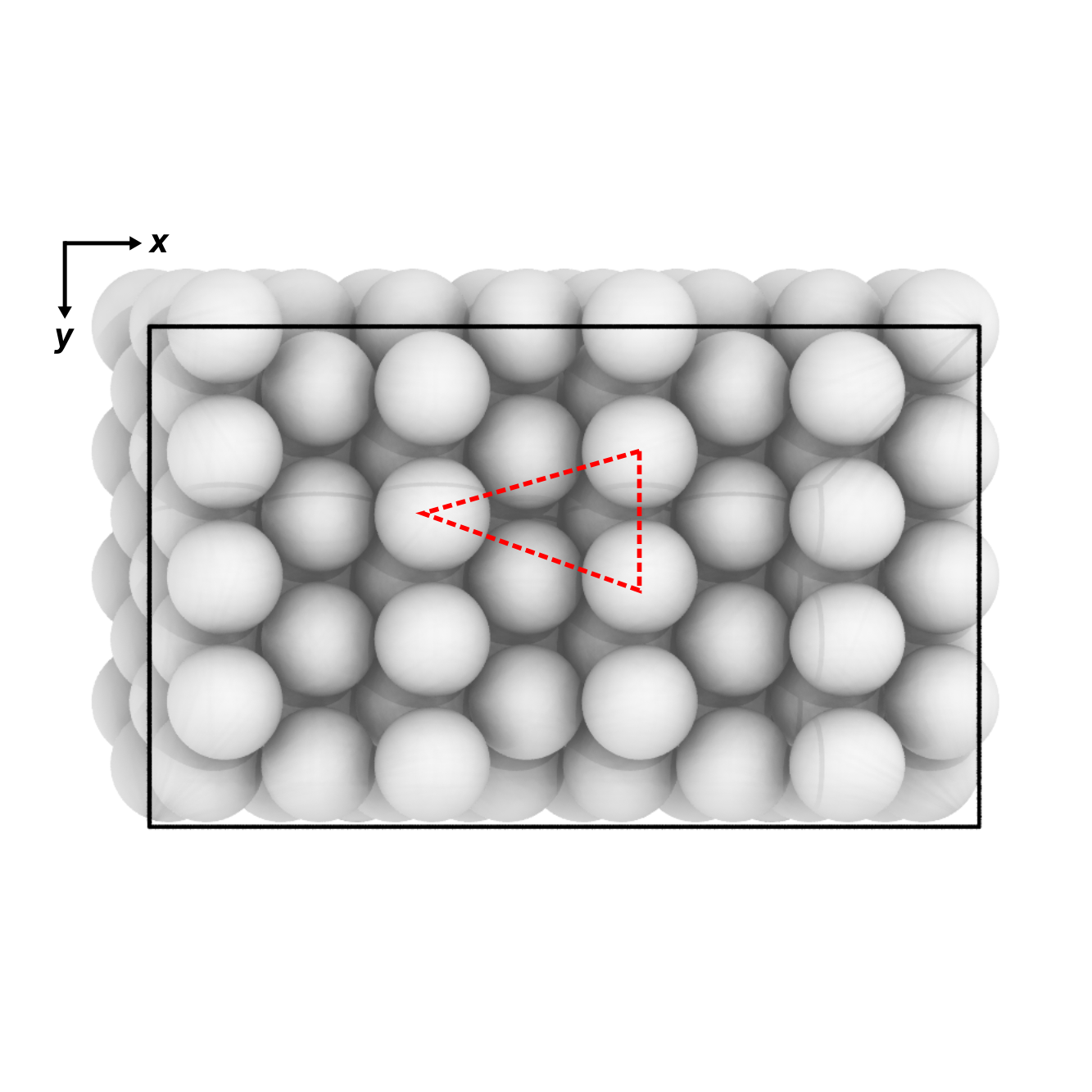}
        \label{fig:top311}
    \end{subfigure}
    \begin{subfigure}[t]{0.24\textwidth}
        \caption{LJ(110) surface top view}
        \includegraphics[width=\textwidth]{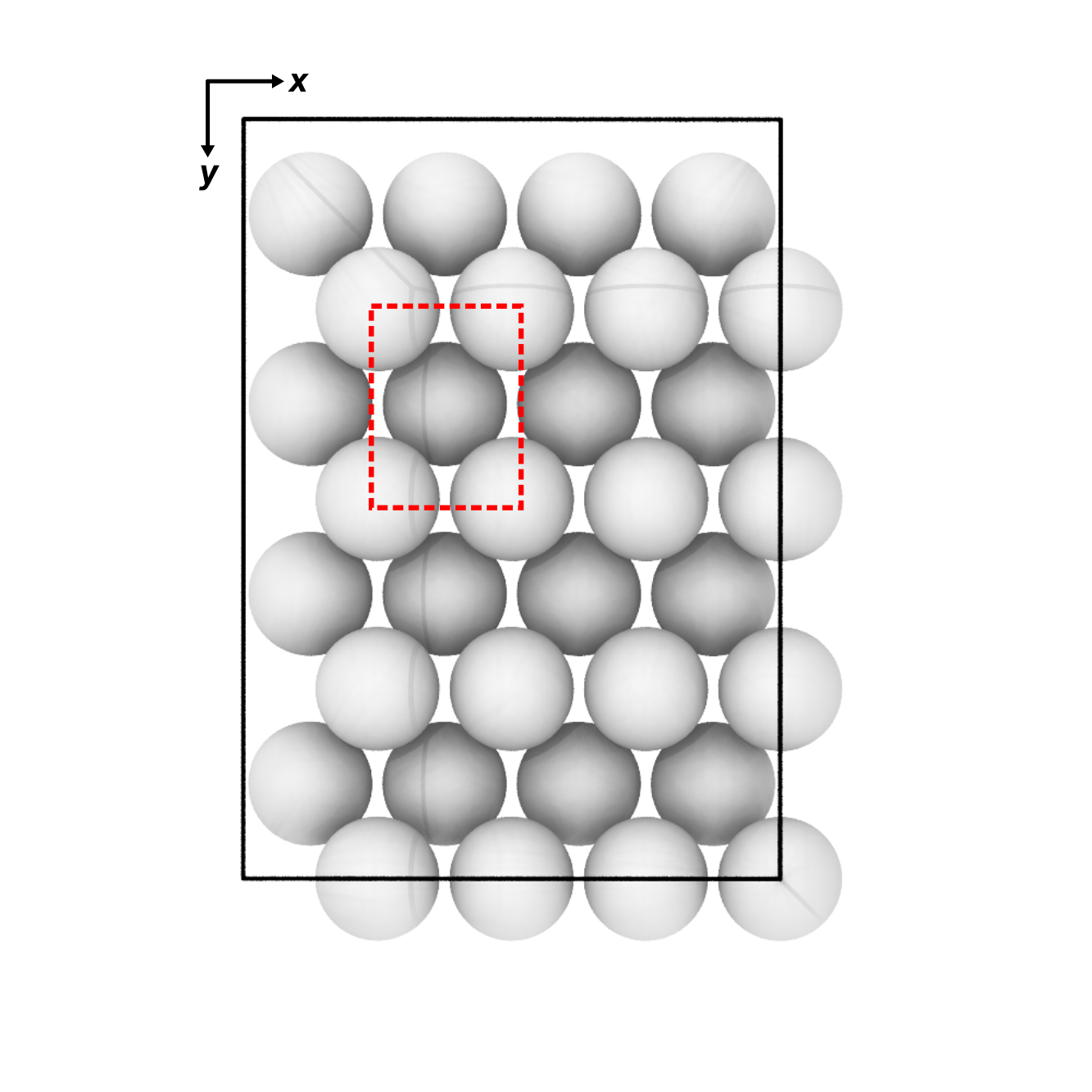}
        \label{fig:top110}
    \end{subfigure}\\
    \begin{subfigure}[t]{0.24\textwidth}
        \caption{LJ(111) surface side view}
        \includegraphics[width=\textwidth]{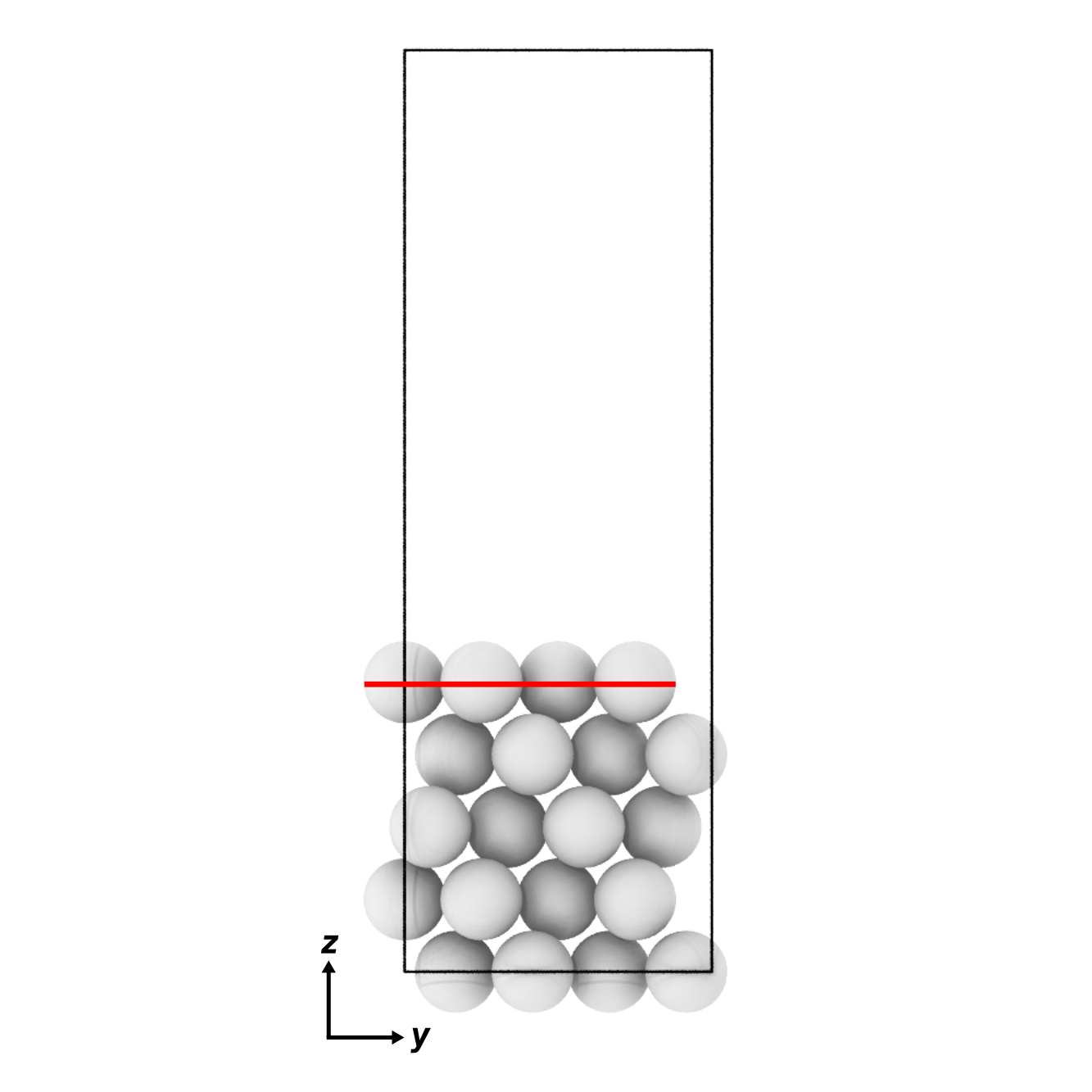}
        \label{fig:side111}
    \end{subfigure}
    \begin{subfigure}[t]{0.24\textwidth}
        \caption{LJ(100) surface side view}
        \includegraphics[width=\textwidth]{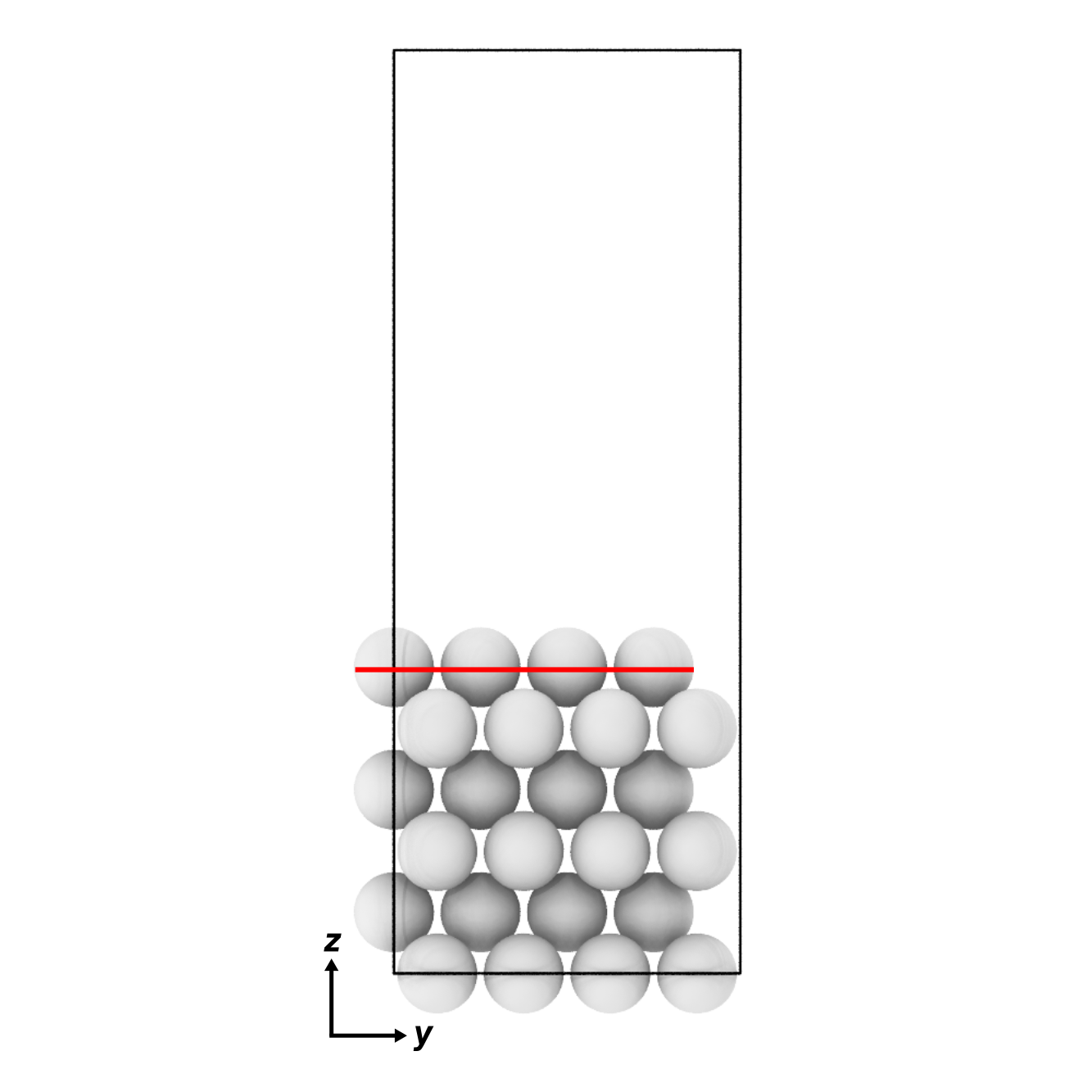}
        \label{fig:side100}
    \end{subfigure}
    \begin{subfigure}[t]{0.24\textwidth}
        \caption{LJ(311) surface side view}
        \includegraphics[width=\textwidth]{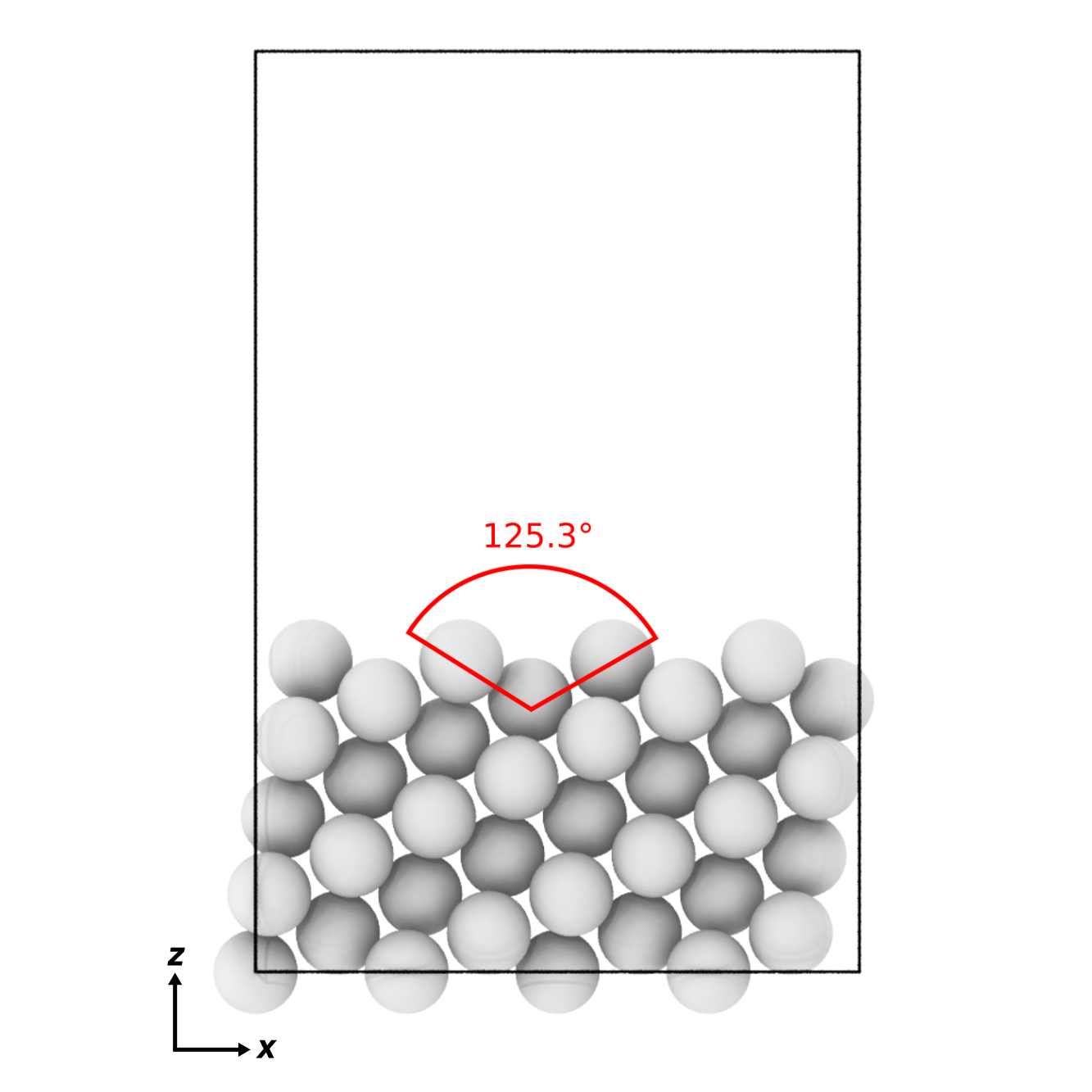}
        \label{fig:side311}
    \end{subfigure}
    \begin{subfigure}[t]{0.24\textwidth}
        \caption{LJ(110) surface side view}
        \includegraphics[width=\textwidth]{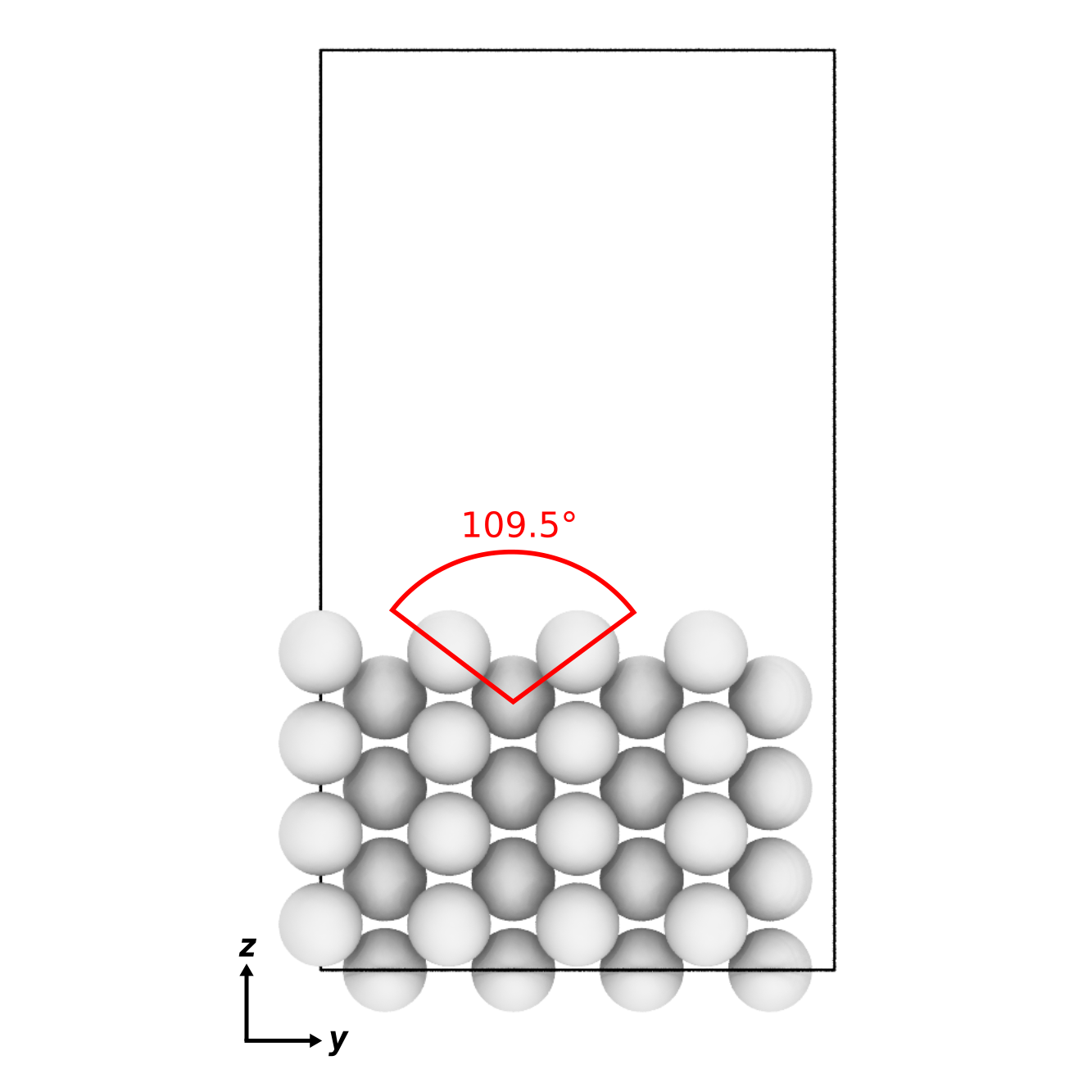}
        \label{fig:side110}
    \end{subfigure}
    \caption{
Top and side views of clean LJ(111), LJ(100), LJ(311), and LJ(110) surfaces. The dashed lines in panels (a)-(d) show a binding site on each surface. The red lines in panels (e) and (f) show the LJ(111) and LJ(100) surfaces, which are considered flat. The angles indicated in red in panels (g) and (h) display the decrease in planarity of the LJ(311) and LJ(110) surfaces, respectively. Note that the angles shown are not the bond angles but the opening of the troughs, viewed from the side. Detailed surface specifications are included in the ESI Table~\ref{tab: SI_facets_params}.
    }
    \label{fig:surf-views}
\end{figure*}

We constructed slabs with a $4\times4$ surface unit cell and the following numbers of layers along the surface normal: five for LJ(111), eight for LJ(110), six for LJ(100), and eight for LJ(311). Different facets require different numbers of layers due to their different interlayer spacing (see Table~\ref{tab: SI_facets_params} in the ESI). We used these numbers of layers to ensure that the bottom surface is the deepest possible layer within the LJ cutoff radius from free particles adsorbed on the surface, representing a semi-infinite bulk crystal beneath the top surface. Figure~\ref{fig:surf-views} displays each facet's top and side views with the surface features highlighted. All four facets explored in this work have unique surface characteristics: LJ(111), a flat surface featuring a triangular lattice-like arrangement with three-fold binding sites; LJ(100), a flat surface featuring a square lattice-like arrangement with four-fold binding sites; LJ(311), a stepped surface characterized by shallow troughs (depth = $0.48\sigma$, opening angle = $125.3^{\circ}$) with elongated triangular binding sites, where an adsorbate can bind three fixed particles on the surface and two additional ones underneath the surface; and LJ(110), another stepped surface, distinguished by its deeper troughs (depth = $0.56\sigma$, opening angle = $109.5^{\circ}$) with stretched square binding sites, where an adsorbate can bind four fixed particles on the surface and an extra one underneath the surface. For each facet, the number of free particles included in the NS calculations ranges from one to 16. With 16 free particles, an ML can be formed on the fixed slab. We analyzed the finite-size effects of our system. Our calculations, which are based on a $4\times4$ surface, effectively capture the fundamental physics of the system. Selecting a $4\times4$ surface guarantees qualitative accuracy while remaining computationally feasible, as Section~\ref{sec: SI-finite-size} of the Electronic Supplementary Information (ESI) explains.

\subsubsection{Nested sampling parameters}

\begin{figure*}[t]
    \centering
    \includegraphics[width=0.8\textwidth]{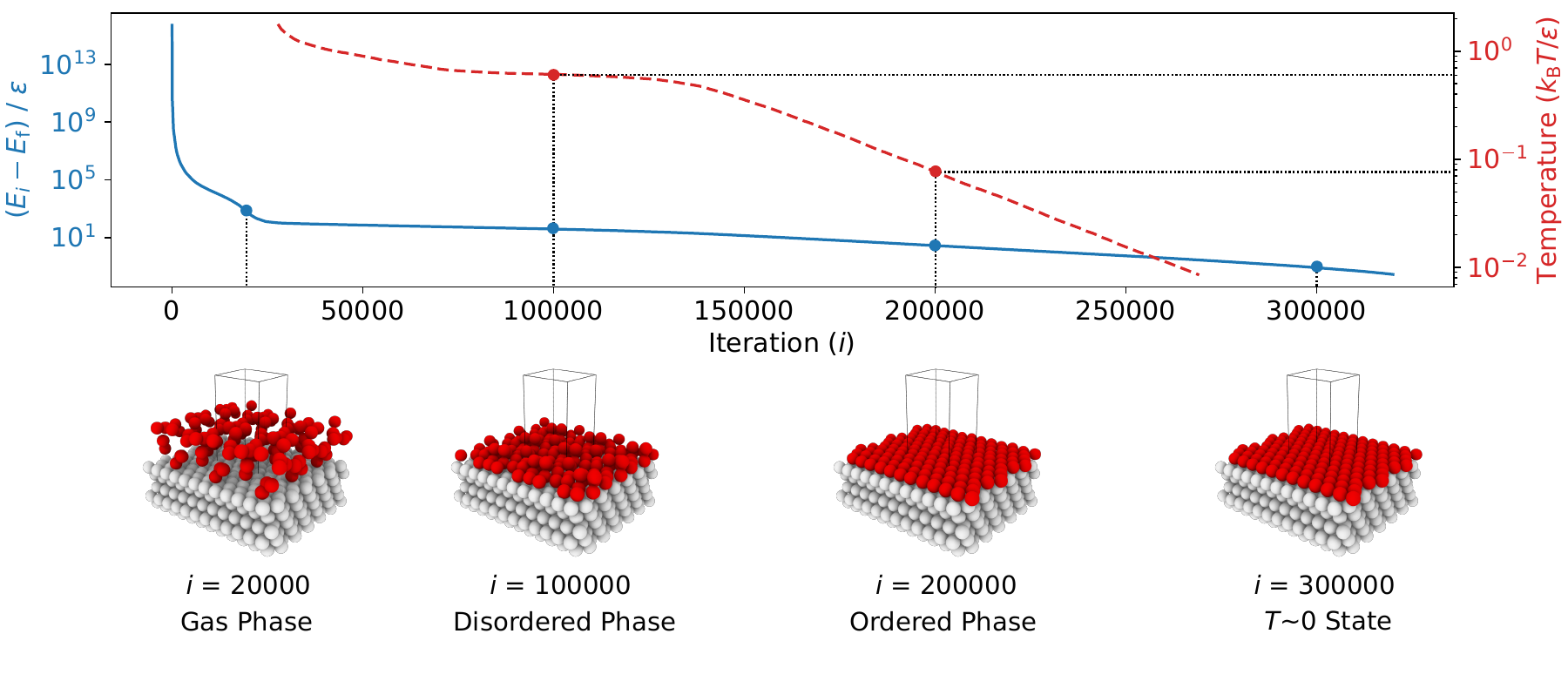}
    \caption{
The recorded potential energies (solid line), $E_i$, relative to the final potential energy (\textit{i.e.}, the lowest potential energy with $T \sim 0$), $E_{\mathrm{f}}$, as well as the estimated temperature (dashed line) within the range between 0.01 and 1.73 $k_{\mathrm{B}} T / \epsilon$, versus the iteration number, $i$, from an NS calculation, are presented using 1,280 walkers at full coverage. The potential energy decreases rapidly during the initial sampling stage (first 30,000 iterations), as the phase space shrinks quickly when the walkers explore the high-temperature configurations in the high-potential-energy region of the PES. Then, the potential energy decreases more slowly, accompanied by a further decrease in temperature towards absolute zero, capturing different configurations with almost degenerate energies. Snapshots of the system illustrate the phase changes from an initial ideal-gas-like phase to a condensed but disordered phase and then to an ordered state. Fixed particles are shown in gray, and free particles in red. Note the logarithmic scale used for energy, and that the energy series is truncated at 320,000 iterations.
    }
    \label{fig:explore_scheme}
\end{figure*}

Our study utilized 80 walkers per free particle for NS on surfaces. The number of NS iterations required increases almost linearly with the number of walkers. We used 250 iterations per walker, as our tests indicated that this number was adequate for achieving the lowest potential energy structures across all coverages. Consequently, for the highest coverage scenario (16 particles per ML), we performed 320,000 NS iterations (calculated as $80\times250\times16$).

The typical variation of potential energy during an NS run is illustrated in Figure~\ref{fig:explore_scheme}. Initially, the algorithm identifies and replaces configurations with high potential energy, which have minimal impact on the partition function due to their small Boltzmann factors. High potential energies result from the proximity between particles, causing increased potential energy due to LJ repulsion. As the NS progresses, the sampled potential energies tend to decrease, eventually converging to the system's lowest potential energy state.

As described in Step~\ref{step2}, the new walkers were generated by cloning an existing walker and then decorrelating it via a 1,600-step random walk while ensuring the energy remained below the current maximum $E_i^{\max}$.
We controlled the acceptance rate of new configurations by incrementally reducing the translational move step size during sampling. We kept the simulation cell's shape and volume constant throughout this process. Every 100 NS iterations, we recorded the configurations being replaced. These recorded structures were later used to calculate surface order parameters and to investigate metastable states.

We performed three independent NS calculations for each system, all with the same NS parameters but starting from different initial walkers. These initial configurations were created by randomly placing particles in \textbf{Region 2} of Figure \ref{fig:surface-setup}. Our implementation of surface NS is available in the open-source \verb|pymatnest| software, accessible at \url{https://github.com/libAtoms/pymatnest}. The input files used to perform the NS calculations with \verb|pymatnest| and all output data are publicly available at DOI: \url{https://doi.org/10.7936/6RXS-103650}.

\section{Results}

We used surface NS to predict the adsorbate phase diagrams for four facets of the face-centered-cubic (fcc) LJ solid. We considered two flat facets, (111) and (100), and two stepped facets, (311) and (110), to analyze the effect of planarity on adsorbate phase equilibria. Additionally, we chose these four specific facets because they are some of the lowest index, lowest surface energy, and highest surface area fraction facets on the Wulff shape of fcc elemental solids. \cite{tran_surface_2016} Our study employs a $4\times4$ surface unit cell for modeling the phase diagrams, which, while computationally efficient, may introduce non-trivial finite-size effects. Given these potential limitations, it is important to exercise caution when extrapolating our findings to real surfaces. However, this research aims to demonstrate surface NS's capabilities within the well-understood LJ system, setting a foundation for more comprehensive and realistic simulations in future studies. Section~\ref{sec: 111} concentrates on the flat LJ(111) surface. We present its phase diagram and examine thermodynamic properties computed from the partition function calculated via NS. A comprehensive analysis of phase transitions on the LJ(111) surface is provided. Section~\ref{sec: 3-surf} elucidates the relationship between phase transitions and surface planarity by comparing the phase diagrams of the LJ(100), LJ(311), and LJ(110) surfaces with that of LJ(111).

\subsection{LJ(111) phase diagram} \label{sec: 111}

\begin{figure*}[p]
    \centering
    \begin{subfigure}[t]{0.45\textwidth}
        \caption{}
        \includegraphics[width=\textwidth]{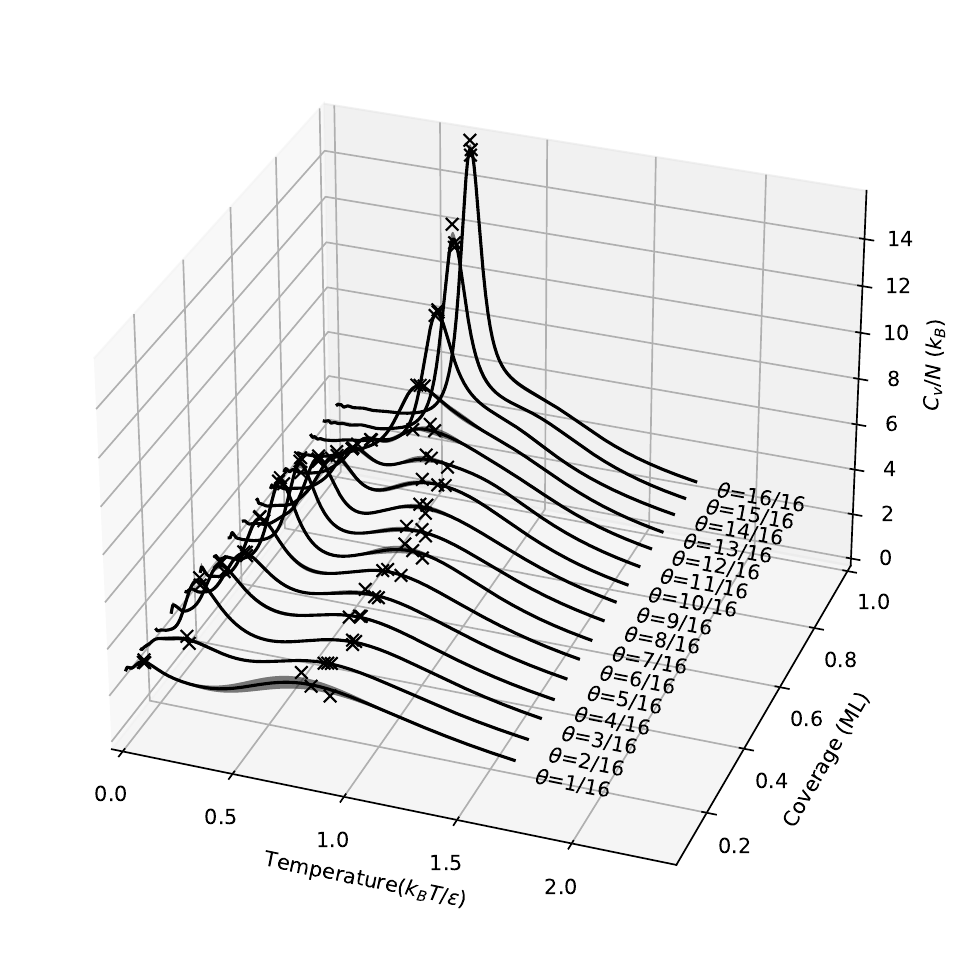}
        \label{fig:surf111_cv}
    \end{subfigure}
    \begin{subfigure}[t]{0.45\textwidth}
        \caption{}
        \includegraphics[width=\textwidth]{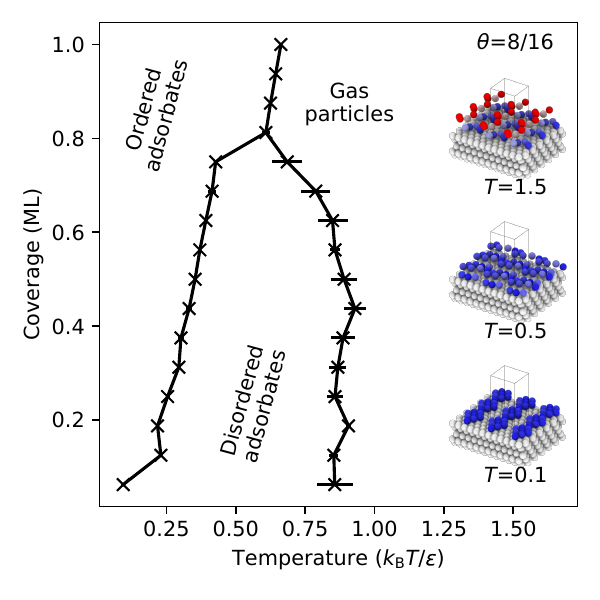}
        \label{fig:surf111_phase_diagram}
    \end{subfigure}\\
    \begin{subfigure}[t]{0.45\textwidth}
        \caption{}
        \includegraphics[width=\textwidth]{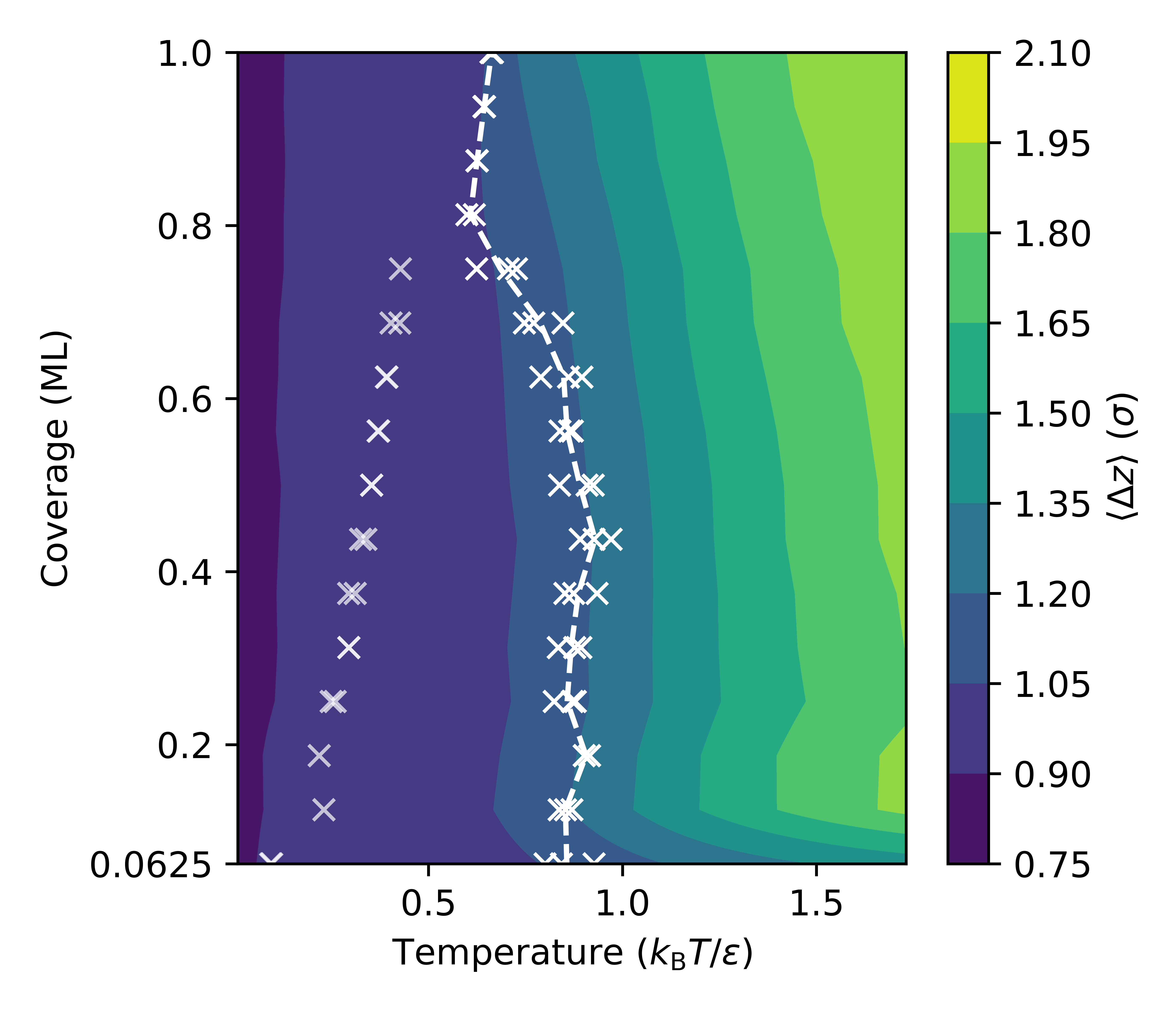}
        \label{fig:surf111_Z}
    \end{subfigure}
    \begin{subfigure}[t]{0.45\textwidth}
        \caption{}
        \includegraphics[width=\textwidth]{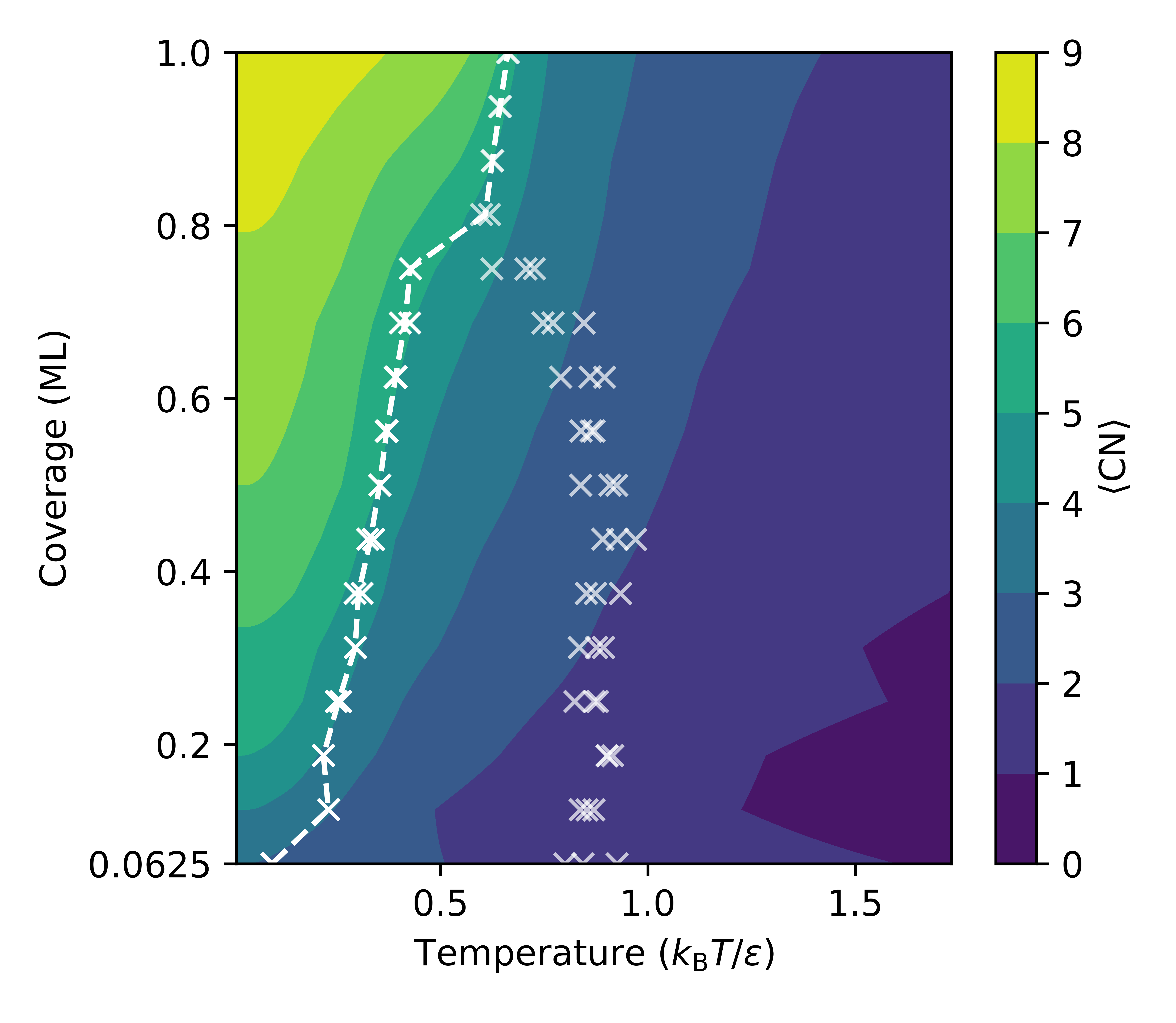}
        \label{fig:surf111_CN}
    \end{subfigure}
    \caption{
Calculated coverage-temperature properties of the flat LJ(111) surface with fractional coverages from $\theta=1/16$ ML up to one ML: (a) heat capacity per free particle, with the peaks on the curves marked with crosses ($\times$), (b) phase diagram with insets showing the maximum-probability structures from each phase at selected temperatures for $\theta=8/16$ ML, where the unit cell is repeated three times in $x$- and $y$-directions for better visibility of the surface structures, (c) average $z$-coordinates of the free particles relative to the topmost layer in the fixed slab. Note that the bulk (111) interlayer spacing is $0.92\sigma$, and (d) free particles' average coordination numbers, including particle-particle and particle-surface bonding. The error bars in panel (b) represent the standard deviations of the peak temperatures for the three independent NS runs at each coverage. The lines between the crosses in panels (b)-(d) are only guides for the eye.
    }
    \label{fig:111-4x4-cv}
\end{figure*}

Following the procedures described in Section~\ref{computational-methods}, we first compute $C_V \left( T^* = k_{\mathrm{B}} T / \epsilon\right)$ for the flat LJ(111) surface with coverages ranging from $\theta=1/16$ ML to one ML (\textit{i.e.}, $\theta=16/16$ ML). The $C_V$ curves for the LJ(111) surface (see Figure~\ref{fig:surf111_cv}) display different behaviors in the lower coverage ($\theta<12/16$ ML) and higher coverage ($\theta\geq12/16$ ML) regimes. In the case of the lower coverage surfaces, two peaks in the $C_V$ curve can be observed. A broad and shallow peak in $C_V$ is noticeable in the lower temperature range between $0.7$ and $1.0~k_{\mathrm{B}} T / \epsilon$. In contrast, at temperatures between $0.2$ and $0.5~k_{\mathrm{B}} T / \epsilon$, a sharper and more distinct peak in $C_V$ emerges. At coverages above $\theta\geq12/16$ ML, the two $C_V$ peaks merge, rising sharply as the coverage increases toward one ML. The resulting coverage-temperature phase diagram for the LJ(111) surface (see Figure~\ref{fig:surf111_phase_diagram}, where each ``$\times$'' marks a peak found by the automatic peak finder) shows two distinct adsorbate phase boundaries for coverages lower than $13/16$ ML, as expected. The two boundaries move towards each other as the coverage increases and eventually meet at a triple-point, at $\theta=13/16$ ML and approximately $0.6~k_{\mathrm{B}} T / \epsilon$.

To characterize the adsorbate phases and their transitions, we calculated two order parameters, $\langle \Delta z \rangle$ (see Figure~\ref{fig:surf111_Z}) and $\langle \mathrm{CN} \rangle$ (see Figure~\ref{fig:surf111_CN}), as described in Section~\ref{sec:phase-equilibria}. As expected, $\langle \Delta z \rangle$ decreases as the temperature decreases because the free particles start adsorbing on the surface; we refer to this process as surface condensation. The $1.2\sigma$ contour coincides with the higher temperature coexistence curve, suggesting that the adsorbed free particles form a quasi-two-dimensional layer with a thickness approximately equal to $1.2\sigma$ upon cooling. Note that $0.92\sigma$ is the interlayer spacing in an LJ(111) bulk, meaning the free particles are now, on average, near the positions where an additional ML should form. Since the higher-temperature adsorbate phase transition corresponds to a vertical ordering of the free particles, the lower-temperature adsorbate phase transition must correspond to the approximately two-dimensional ordering of the free particles (or adsorbates, after condensation) within the surface layer. To quantify this surface-layer ordering, we calculated $\langle \mathrm{CN} \rangle$, which increases with decreasing temperature (see Figure~\ref{fig:surf111_CN}). At high temperatures and low coverages, the free particles rarely interact with one another, resulting in a $\langle \mathrm{CN} \rangle \approx 0$. However, once the temperature is less than that of the condensation, $\langle \mathrm{CN} \rangle$ quickly reaches its maximum possible value: for example, three for one free particle (because the free particle occupies a hollow site between three fixed surface particles), and four for two free particles (because the two free particles occupy neighboring hollow sites). The maximum $\langle \mathrm{CN} \rangle$ is nine, where an adsorbed free particle is close-packed by six adsorbed free particles at neighboring hollow sites. Overall, the lower temperature coexistence curve coincides with a contour in $\langle \mathrm{CN} \rangle$, separating a lower coordination adsorbate phase with disordered adsorbates above the transition temperature from a higher coordination adsorbate phase with ordered adsorbates below the transition temperature.

Interestingly, the $4\times4$ (and $6\times6$, see Section~\ref{sec: SI-finite-size}) LJ(111) surface has a triple point near a coverage of three-quarters ($\theta=13/16$ ML) and at a temperature between the two adsorbate phase transitions observed for lower coverages. For coverages $\theta\geq13/16$, the $C_V$ curves in Figure~\ref{fig:surf111_cv} show only one sharp peak. Given the lack of a lower coordination adsorbate phase with disordered adsorbates at intermediate temperatures for coverages $\theta\geq13/16$ ML, we will refer to this process as surface deposition, where gas-phase free particles form an adsorbate phase with ordered adsorbates below the deposition temperature. One can see from Figure~\ref{fig:surf111_Z} that the phase boundary now coincides with the $\langle \Delta z \rangle=1.05\sigma$ contour versus the $\langle \Delta z \rangle=1.20\sigma$ contour for the lower-coverage, higher-temperature transition, showing that the phase transition processes happen closer to the fixed slab compared to the surface condensation. We can rationalize the origin of these two different behaviors for lower and higher coverages -- \textit{i.e.}, condensation and deposition, respectively -- by comparing the stable surface structures below the condensation and deposition temperatures, respectively (see Section~\ref{sec: SI-structures} in the ESI). For lower coverages, the free particles form islands and stripes (see the inset in Figure~\ref{fig:surf111_phase_diagram} for an example structure at $\theta=8/16$ ML and $T=0.1~k_{\mathrm{B}} T / \epsilon$) on the surface, with many isoenergetic options for reconfiguration. However, for higher coverages, the free particles form a continuous ML with vacancies, which have limited options for reconfiguration.

Another observation from our NS results is that, on the LJ(111) surface, the adsorbed free particles form a hexagonal-close-packed (hcp) ML on top of the fcc LJ solid at absolute zero. As shown in Reference \citenum{LJPolytypism}, the LJ solid's lowest potential energy state stacking structure depends on how the potential around the truncation distance is treated. In our specific setup, the potential energy of the hcp ML is less than that of the fcc ML by $2\times10^{-7}\epsilon$ per adsorbed free particle. Thus, NS finds the hcp ML the lowest potential energy structure. On the other hand, molecular dynamics simulations show that crystal growth on an LJ(111) surface consistently forms a mix of fcc and hcp structures, with the fcc LJ(111) surface being slightly kinetically favored. \cite{baez_kinetics_1995, somasi_computer_2001} However, Somasi \textit{et al.} used an LJ cutoff radius of $2.5 \sigma$; therefore, based on Reference \citenum{LJPolytypism}, we cannot simply compare their system to ours.

\subsection{Effect of surface geometry on adsorbate phase diagrams} \label{sec: 3-surf}

The preceding section deepened our understanding of phase transitions on the LJ(111) surface. This section expands our investigation to three additional facets of the LJ solid, demonstrating the versatility of our NS approach in studying adsorbate phase diagrams. The heat capacities for the LJ(100), LJ(311), and LJ(110) surfaces are detailed in Figures \ref{fig:cv100}, \ref{fig:cv311}, and \ref{fig:cv110}, respectively. Correspondingly, the phase diagrams for these surfaces are presented in Figures \ref{fig:phase100}, \ref{fig:phase311}, and \ref{fig:phase110}.

\subsubsection{Lattice type}

\begin{figure*}[htb]
    \centering
    \begin{subfigure}[t]{0.33\textwidth}
        \caption{LJ(100) heat capacity}
        \includegraphics[width=\textwidth]{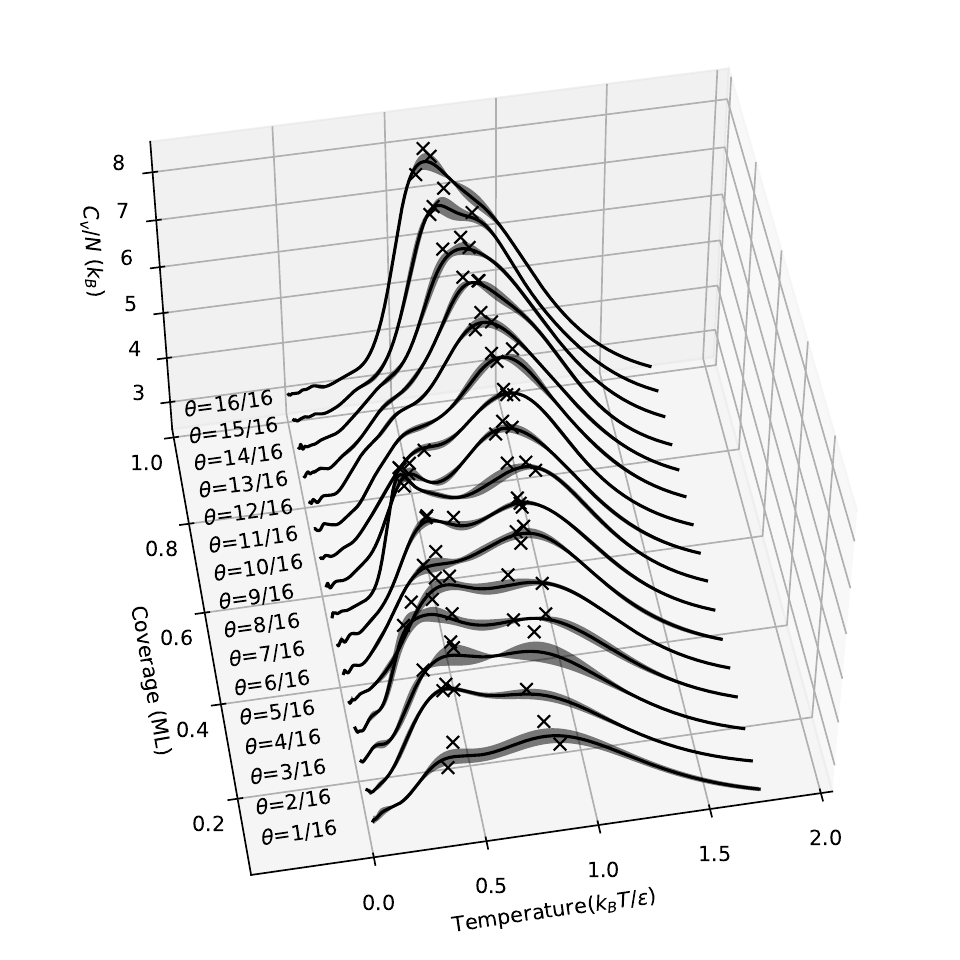}
        \label{fig:cv100}
    \end{subfigure}
    \begin{subfigure}[t]{0.33\textwidth}
        \caption{LJ(311) heat capacity}
        \includegraphics[width=\textwidth]{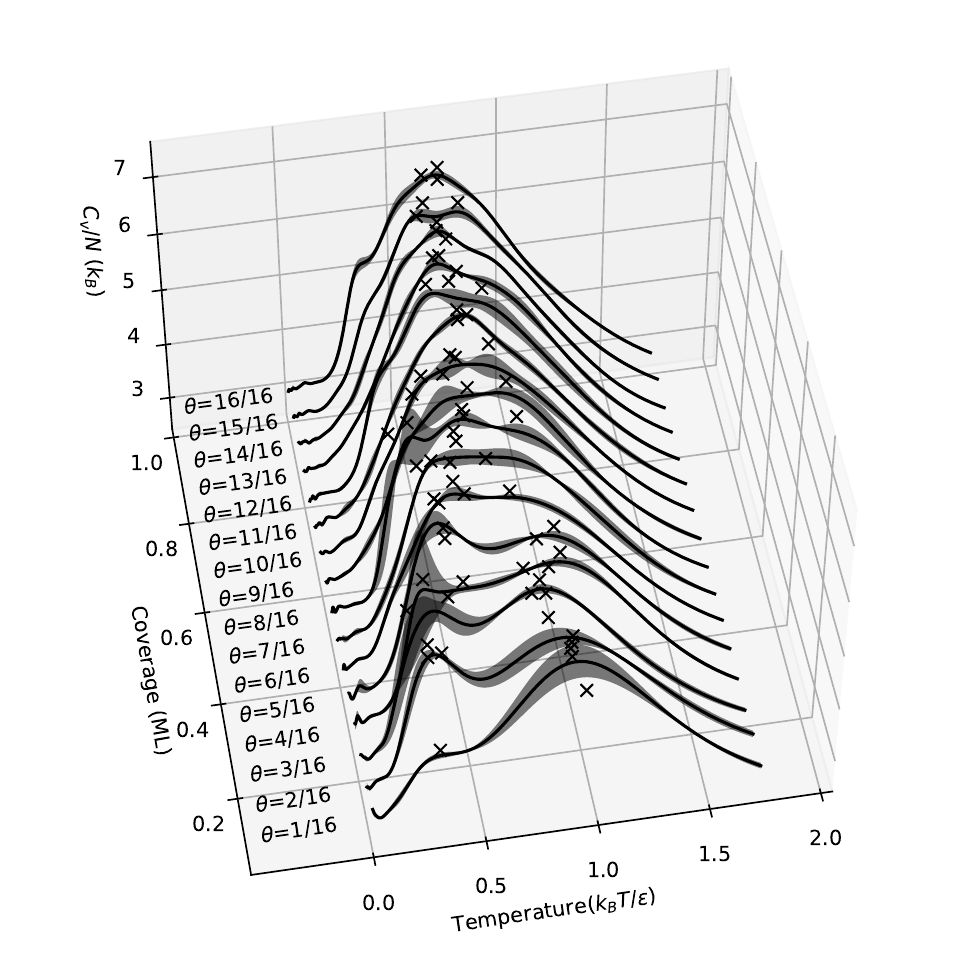}
        \label{fig:cv311}
    \end{subfigure}
    \begin{subfigure}[t]{0.33\textwidth}
        \caption{LJ(110) heat capacity}
        \includegraphics[width=\textwidth]{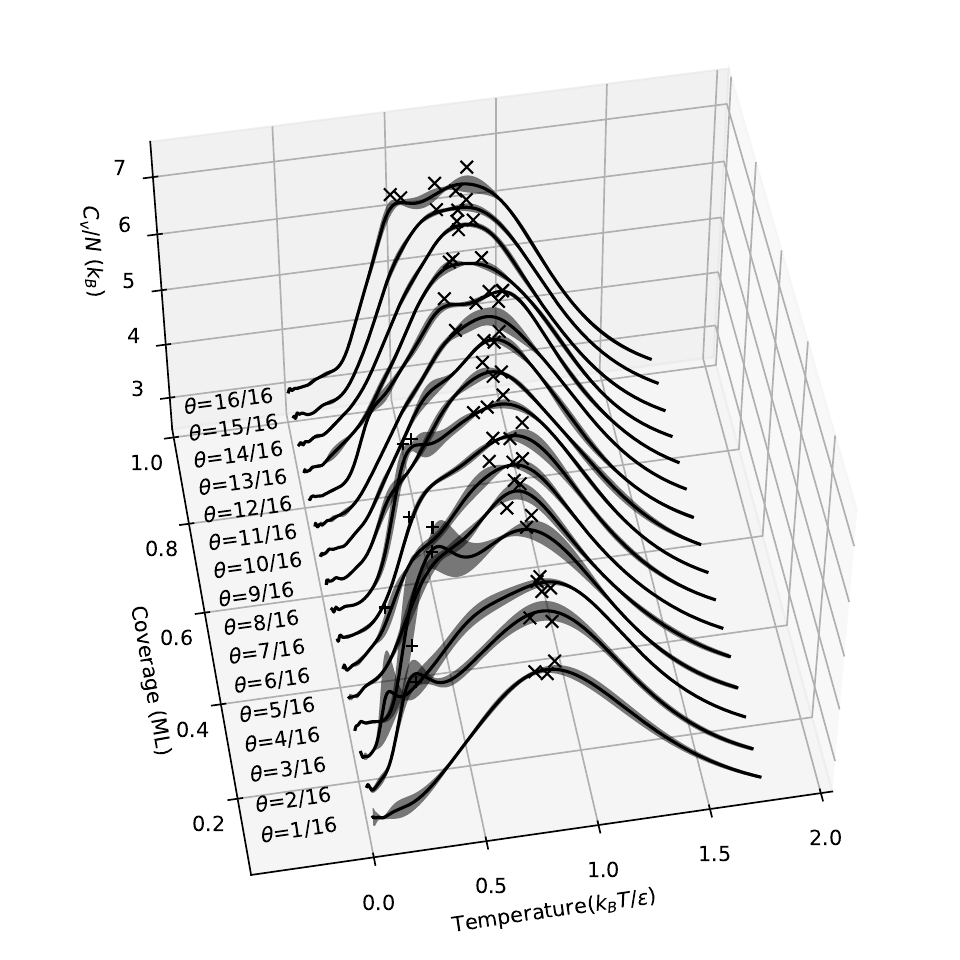}
        \label{fig:cv110}
    \end{subfigure}\\
    \begin{subfigure}[t]{0.33\textwidth}
        \caption{LJ(100) phase diagram}
        \includegraphics[width=\textwidth]{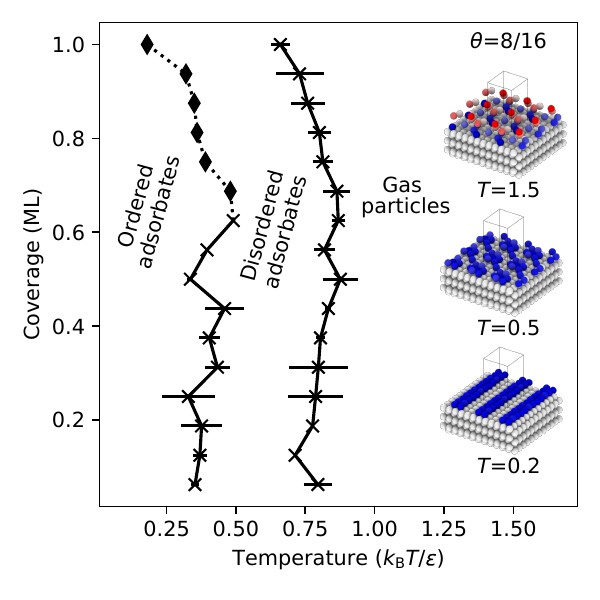}
        \label{fig:phase100}
    \end{subfigure}
    \begin{subfigure}[t]{0.33\textwidth}
        \caption{LJ(311) phase diagram}
        \includegraphics[width=\textwidth]{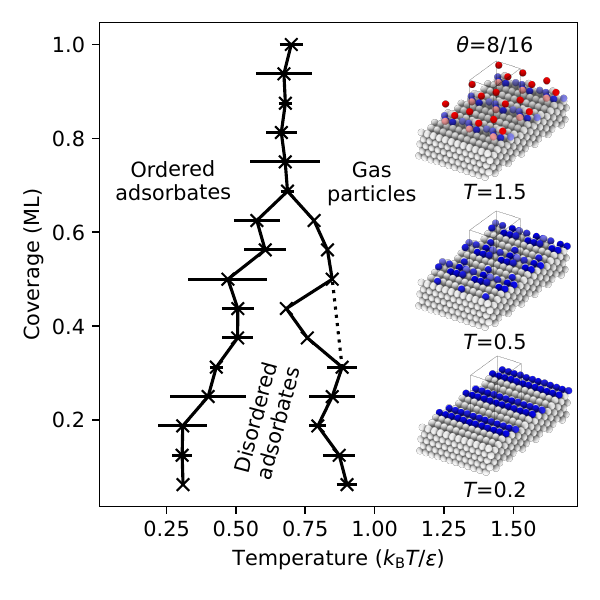}
        \label{fig:phase311}
    \end{subfigure}
    \begin{subfigure}[t]{0.33\textwidth}
        \caption{LJ(110) phase diagram}
        \includegraphics[width=\textwidth]{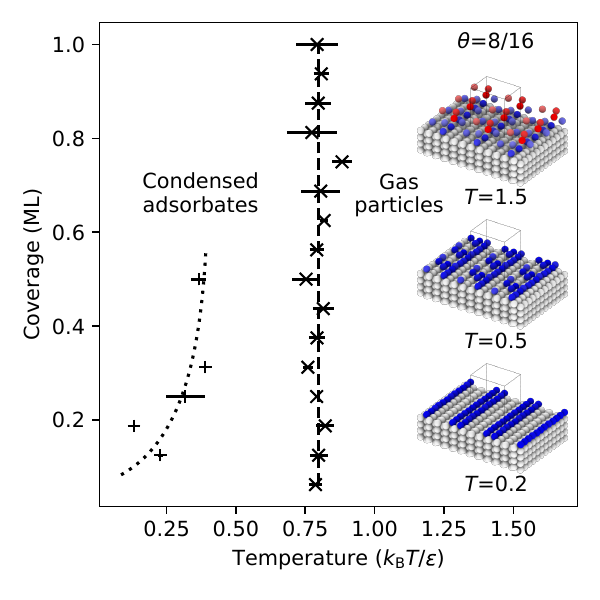}
        \label{fig:phase110}
    \end{subfigure}
    \caption{
Calculated heat capacities and coverage-temperature phase diagrams for three facets of an LJ solid with decreasing planarity: LJ(100), LJ(311), and LJ(110), with the peaks on the curves marked with crosses ($\times$). The plus ($+$) markers in panels (c) and (f) indicate that irregular side peaks occur on LJ(110) at low coverages with no clear trend. The lines between the markers in panels (d)-(f) are only guides for the eye, and the dotted lines are speculative. The insets show the maximum probability structures from each phase at selected temperatures for $\theta=8/16$ ML, where the unit cell is repeated three times in $x$- and $y$-directions for better visibility of the surface structures. The diamond markers ($\blacklozenge$) indicate the disappearing peaks not found by the automated procedure.
    }
    \label{fig: phases}
\end{figure*}

First, we contrast the LJ(100) and LJ(111) surfaces, each flat but differing in the two-dimensional lattice types of the surface particles: a square lattice for LJ(100) and an equilateral triangular lattice for LJ(111). The phase diagrams show that condensation transitions on LJ(100) occur at temperatures similar to those on LJ(111). The ordering process, as established in LJ(111), is also applicable to LJ(100), as evidenced by our analysis of the heat capacity peaks, as well as the order parameters (refer to Figures~\ref{fig:surf100_Z} and \ref{fig:surf100_CN} in the ESI). Note that the disordered-to-ordered phase boundary at $\theta>10/16$ ML, indicated by the $\blacklozenge$ markers in Fig.~\ref{fig:phase100}, is determined manually by locating the shoulder peaks on the $C_V$ curves. LJ(100) lacks the triple point evident in LJ(111), as the surface condensation and disordered-to-ordered phase boundaries do not converge. The progression of the disordered-to-ordered phase boundary in our study is similar to the trends observed in grand canonical MC simulations of a two-dimensional LJ square lattice. Notably, the transition temperature range we identified, $T = 0.2 - 0.5~k_{\mathrm{B}} T / \epsilon$ (see the lower temperature transition in Figure~\ref{fig:phase100}), aligns well with the findings reported in Patrykiejew and Borowski (see Figure~4 in Reference~\citenum{patrykiejew_two-dimensional_1991}). This correspondence underscores the relevance and applicability of lattice models in understanding adsorbate phase behavior. \cite{stampfl_first-principles_1999, mize_insight_2022}

\subsubsection{Planarity}

This part examines the complexities of stepped surfaces: LJ(311) and LJ(110). Unlike flat, two-dimensional systems, these surfaces present unique challenges for traditional lattice models due to their intricate geometries. The stepped surfaces are characterized by washboard-like structures with troughs, or ``missing rows,'' extending along one unit cell dimension. In contrast, the cell is elongated in the perpendicular dimension, markedly increasing the surface area. Due to their unique, reduced-symmetry geometries, these surfaces feature significantly non-degenerate binding sites.

Moreover, redefining what constitutes an ML on these stepped surfaces is important. In this context, an ML consists of all particles at the same height, differing from flat surfaces where all exposed particles are typically considered part of an ML. This approach effectively redefines the ML, preventing the misleading impression of double particle counting per ML. To facilitate a direct comparison of phase diagrams, we maintained the same number of free particles (and, consequently, degrees of freedom) as used for the LJ(111) and LJ(100) surfaces. This consistency is key to understanding the nuanced behaviors of these complex surfaces.

In our examination of the LJ(311) surface, the phase diagram, shown in Figure~\ref{fig:phase311}, bears a resemblance to that of LJ(111), featuring two adsorbate phase transitions at low coverages and a triple point around three-quarters coverage. However, a key distinction is the lack of a pronounced increase in heat capacity at higher coverages above the triple point. Additionally, on LJ(311), high-temperature phase transitions linked to the condensation process become more dominant, overshadowing the lower-temperature ordering transition and complicating phase boundary delineation. This effect is further amplified on the LJ(110) surface, as Figure~\ref{fig:phase110} shows. Here, the low-temperature peaks in the $C_V$ curves are overshadowed by broad, high-temperature peaks, making the phase boundary between the ordered and disordered phases challenging to identify. Consequently, we focus only on delineating the phase boundary between the gas and condensed phases for LJ(110), where the condensation process is markedly more complex and distinct from flat surfaces. Interestingly, condensation on LJ(110) appears to be coverage-independent, as indicated by the vertical dashed line in Figure~\ref{fig:phase110}, representing a consistent phase transition temperature $\overline{T}_\theta = 0.8~k_{\mathrm{B}} T / \epsilon$ across all coverages.

The stepped nature of LJ(311) and LJ(110) surfaces, which have deeper and more directional binding sites in the troughs, drives phase transitions primarily through enthalpy changes, contrasting with the entropy-driven ordering transitions on flat surfaces. The reduction in surface symmetry on these stepped surfaces significantly diminishes entropy contributions, as free particles are more constrained by the stronger binding and the reduced number of energetically equivalent adsorption sites.

\begin{figure}[t]
    \centering
    \includegraphics[width=0.5\textwidth]{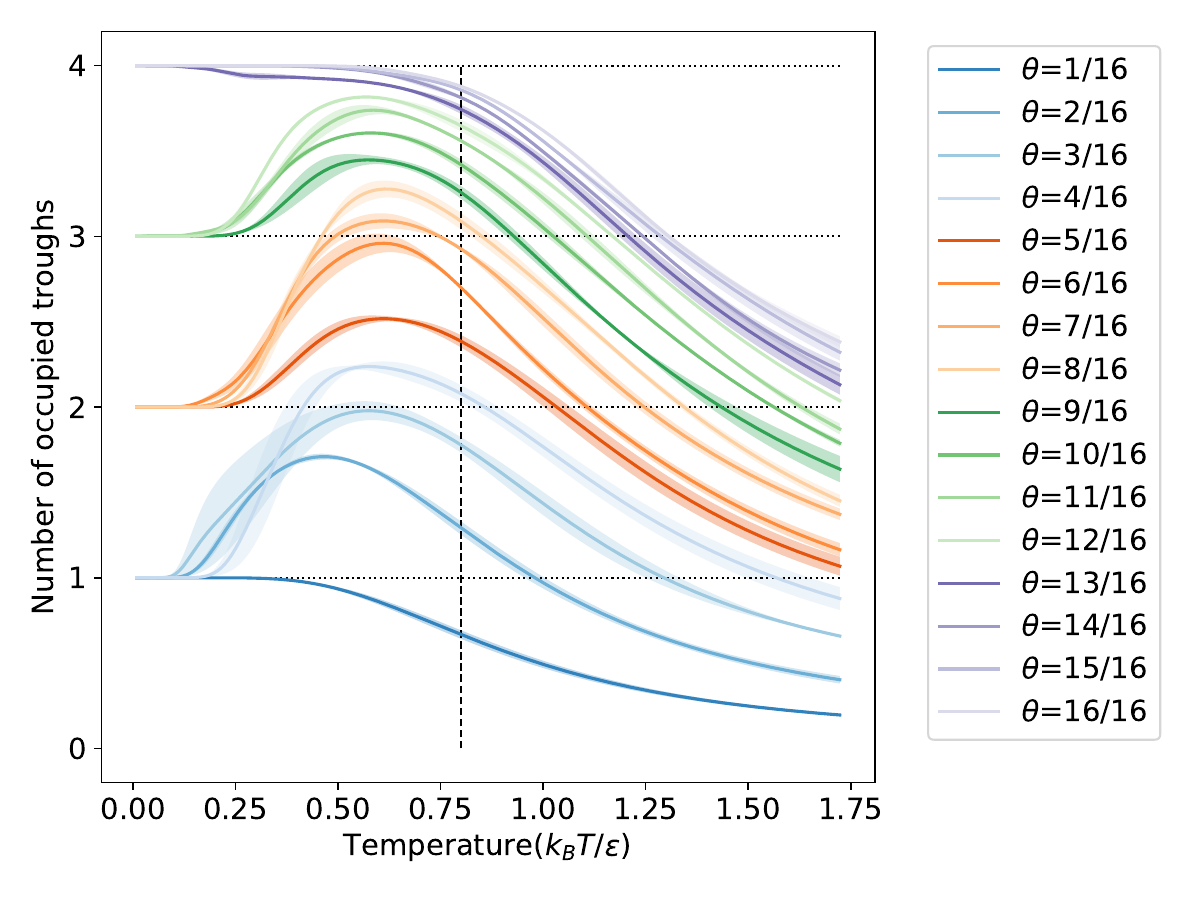}
    \caption{
The ensemble-average number of occupied troughs on LJ(110) as a function of temperature at each coverage is shown. The shaded area around each curve indicates the standard deviation in the number of occupied troughs for the three independent NS runs. The horizontal dotted lines show the integer number of troughs. The vertical dashed line indicates the phase transition temperature, $\overline{T}_\theta = 0.8~k_{\mathrm{B}} T / \epsilon$.
    }
    \label{fig:110-trough-number}
\end{figure}

To further understand the microscopic aspects of phase transitions, we computed the ensemble average number of occupied troughs on the LJ(110) surface at various temperatures and coverages, as depicted in Figure~\ref{fig:110-trough-number}. Observing the temperature range from high to low in Figure~\ref{fig:110-trough-number}, we notice an increasing trend in trough occupancy as free particles begin to populate the troughs, continuing until the temperature reaches $\overline{T}_\theta$ for all coverages. Below $\overline{T}_\theta$, for coverages greater than 12/16 ML, the number of occupied troughs monotonically approaches four, the total count in the unit cell. For lower coverages ($\leq 12/16$ ML), the occupied trough count trends towards $\lceil N/4 \rceil$, reflecting the ability of each trough to house up to four particles in an ML configuration. This behavior is further evidenced by the rapid rearrangement of free particles into the same troughs to maximize coordination, as shown in Figure~\ref{fig:surf110_CN} in the ESI. Such rearrangements, typically accompanied by a reduction in entropy, result in a more ordered state within the same trough, particularly noticeable at lower coverages with numerous unoccupied binding sites. The observed ordering process is linked to the shoulder peaks on the $C_{\mathrm{V}}$ curves in Figure~\ref{fig:cv110} at specific coverages. Our order parameter analysis indicates that these peaks signify an ordering phase transition occurring near the temperatures of the condensation transition.

\section{Discussion}

\subsection{Surface planarity and adsorbate disorder}

We first address the disappearance of the disordered adsorbate phase when transitioning from flat to stepped surfaces. Disordered adsorbate phases are observed at intermediate temperatures on the flat LJ(111) and LJ(100) surfaces. However, this is not true for the stepped LJ(110) surface. As Section~\ref{sec: 3-surf} outlines, the LJ(110) surface transitions directly from gas to ordered adsorbates during condensation. This directness suggests a close link between the ordering process and condensation, influenced by the surface geometry. On the LJ(110) surface, adsorbates settle deeply into troughs post-condensation for maximum coordination, merging the condensation and ordering processes. This overlap is evident from concurrent peaks in heat capacities, indicating a need for refined analytical approaches to separate these overlapping peaks. Such challenges are common in numerical simulations and experimental studies (e.g., References \citenum{migone_melting_1984} and \citenum{butler_completion_1980}), emphasizing the crucial role of surface topology in adsorption and ordering behaviors.

Furthermore, the LJ(311) surface, with its shallower and wider troughs (depth = $0.48\sigma$, opening angle = $125.3^{\circ}$), compared to LJ(110) (depth = $0.56\sigma$, opening angle = $109.5^{\circ}$), exhibits a rougher topology than flat surfaces. This is reflected in the LJ(311) phase diagram, which shows a narrower range of temperatures and coverages where the disordered adsorbate phase is stable (see Figure~\ref{fig:phase311}). This phase diagram is a transitional point between flat and stepped surfaces, highlighting the critical role of planarity in determining the equilibrium structures of adsorbates on LJ solids.

The troughs of stepped surfaces reduce the number of isoenergetic adsorption sites, decreasing the configurational entropy of adsorbed particles compared to flat surfaces. As a result, the boundary of the entropy-stabilized disordered adsorbate phase is less distinct on LJ(110) and at high coverages on LJ(311). One key observation is the correlation between the presence of a triple point on the phase diagram and surface geometry. Despite their different topologies, the triple point is evident on LJ(111) and LJ(311) surfaces, which feature triangular binding sites. In contrast, the LJ(100) and LJ(110), with their square and rectangular binding sites, surfaces show no clear triple point. Instead, they display pronounced coverage-independent phase transitions.

\subsection{Transferability of surface NS}

This work demonstrates that NS can compute the coverage-temperature phase diagrams of solid surfaces. However, other state variables, such as pressure and chemical potential, can also be considered. For instance, NS could be performed with particle addition and removal steps and translational moves to construct composition-temperature phase diagrams in the case of semi-grand canonical NS \cite{rosenbrock_machine-learned_2021} and chemical potential-temperature phase diagrams in the case of grand canonical NS. Developing the latter will be essential to predict surface reconstructions under operating conditions, \textit{i.e.}, \textit{operando} chemical potentials, robustly. Our NS implementation is also not limited to studying surface-adsorbate phase equilibria. We define a general interfacial system with two interacting subsystems: (1) a fixed host phase, which in our case is a surface slab model (see Figure~\ref{fig:surface-setup}) and (2) a free guest phase, which in our case are particles that start as a randomly and uniformly distributed gas and end as an adsorbed monolayer. The generalization of this setup to other interfacial systems involves substituting the host and guest with subsystems of interest. For example, consider the solid surface-liquid solvent interface in heterogeneous catalysis and typical rechargeable batteries. Here, the host would be the solid catalyst or electrode surface, and the guest would be the liquid phase. The liquid phase could be treated using explicit solvent particles coupled with implicit solvation in a continuous dielectric medium to improve computational efficiency. Alternatively, our approach could be extended to study interfaces between two solids, such as those at grain boundaries (where reconstructions called complexions can form) and electrical junctions. In these cases, the host and guest would be solids, but care would have to be taken in selecting or designing sampling moves to increase the acceptance ratio, as it can be low in condensed phases.

\subsection{Opportunities for improving surface NS}

Finally, we have adopted a simplified view of the surface, \textit{i.e.}, as a host whose constituent particles can interact with other particles but cannot move for computational convenience and model simplicity. While this simplification allows us to focus exclusively on the free particle-surface interplay, it precludes the surface from contributing to the free energy via its vibrational and configurational degrees of freedom, and it neglects effects such as adsorbate-induced surface reconstructions. Such processes can be critical for modeling specific systems realistically, for example, CO adsorption on Pt(100)\cite{kinomoto_infrared_1991,rasko_co-induced_2003}, or hydrogen adsorption on W(110)\cite{hulpke_hydrogen-induced_1992,altshuler_vibrational_1997}. To include at least some of these surface contributions, we propose the introduction of ``flexible'' surface particles that are neither fixed nor free but confined harmonically to their lattice sites. Such an approach, which we intend to develop in future work, would allow NS to capture the effect of harmonic surface vibrations in the system partition function. In future work, we also aim to benchmark the accuracy and efficiency of NS against methods such as numerical integration, replica exchange,\cite{swendsen_replica_1986, marinari_simulated_1992, sugita_replica-exchange_1999, shirts_statistically_2008} and the Wang-Landau algorithm.\cite{wang_efficient_2001, morozov_accuracy_2007} This comparison will provide valuable insights into the strengths and limitations of each approach.

\section{Conclusions}

In this study, we developed the nested sampling (NS) algorithm for surfaces, extending its use to predict coverage-temperature adsorbate phase diagrams and compute surface thermodynamic properties at finite temperatures. We employed surface NS to construct partition functions for Lennard-Jones (LJ) solid surfaces with fixed and free particles. We calculated the constant-volume heat capacity from these partition functions, using its peaks to delineate coverage-temperature adsorbate phase diagrams.

Our analysis revealed that free particles on these surfaces typically undergo two phase transitions: a higher-temperature, enthalpy-driven condensation followed by a lower-temperature, entropy-driven reordering. This NS-based approach effectively resolved phase diagrams for both flat and stepped surfaces. Order parameters were crucial for stepped surfaces, where phase boundaries are less clear. These parameters, calculated as ensemble averages of observables from the partition function, provide statistical insights into complex surface behaviors.

This work enhances our understanding of surface processes and paves the way for future implementations of NS on open thermodynamic systems and multi-species surfaces. Such advancements are critical for identifying interfacial phases that are key in material performance for commercial, industrial, and climate change mitigation applications.


\section*{Data availability}
The data supporting this study's findings are publicly available at WashU Research Data DOI: \url{https://doi.org/10.7936/6RXS-103650}. The open-source package \verb|pymatnest| is freely available on GitHub at \url{https://github.com/libAtoms/pymatnest}.

\section*{Conflicts of interest}
There are no conflicts to declare.

\section*{Acknowledgments}
RBW acknowledges support from the National Science Foundation under Grant No.~2305155. LBP acknowledges support from the EPSRC through an individual Early Career Fellowship (EP/T000163/1). MY used resources from the Argonne Leadership Computing Facility, a US Department of Energy Office of Science user facility at Argonne National Laboratory. LBP used computing facilities provided by the Scientific Computing Research Technology Platform of the University of Warwick.



\balance


\bibliography{Surfaces} 
\bibliographystyle{rsc} 



\clearpage 

\renewcommand\thefigure{S\arabic{figure}}
\setcounter{figure}{0} 
\renewcommand\thetable{S\arabic{table}}
\setcounter{table}{0} 
\renewcommand{\thesubsection}{S\arabic{subsection}}


\section*{Electronic Supplementary Information}

\subsection{Finite-size effects\label{sec: SI-finite-size}}

\begin{figure*}[htbp]
    \centering
    \begin{subfigure}[t]{0.43\textwidth}
        \caption{}
        \includegraphics[width=\textwidth]{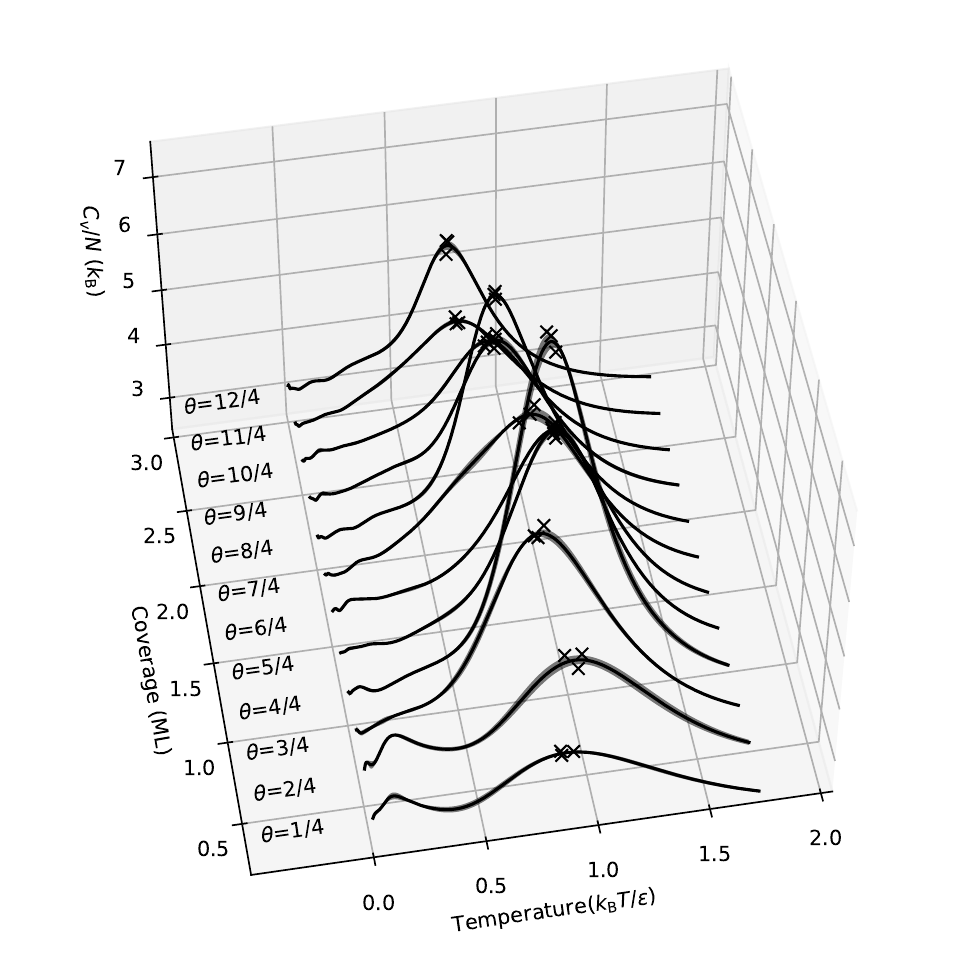}
        \label{fig:surf111_2x2_cv}
    \end{subfigure}
    \begin{subfigure}[t]{0.43\textwidth}
        \caption{}
        \includegraphics[width=\textwidth]{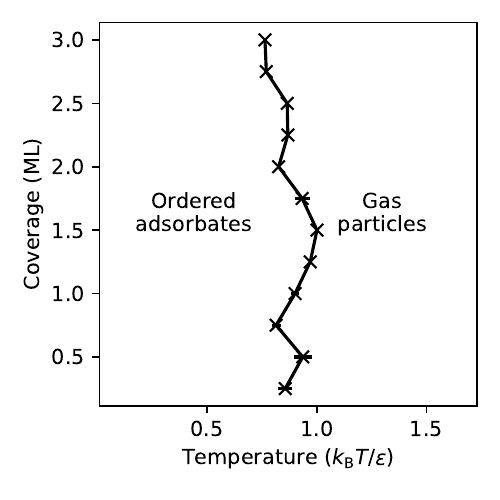}
        \label{fig:surf111_2x2_phase_diagram}
    \end{subfigure}
    \begin{subfigure}[t]{0.43\textwidth}
        \caption{}
        \includegraphics[width=\textwidth]{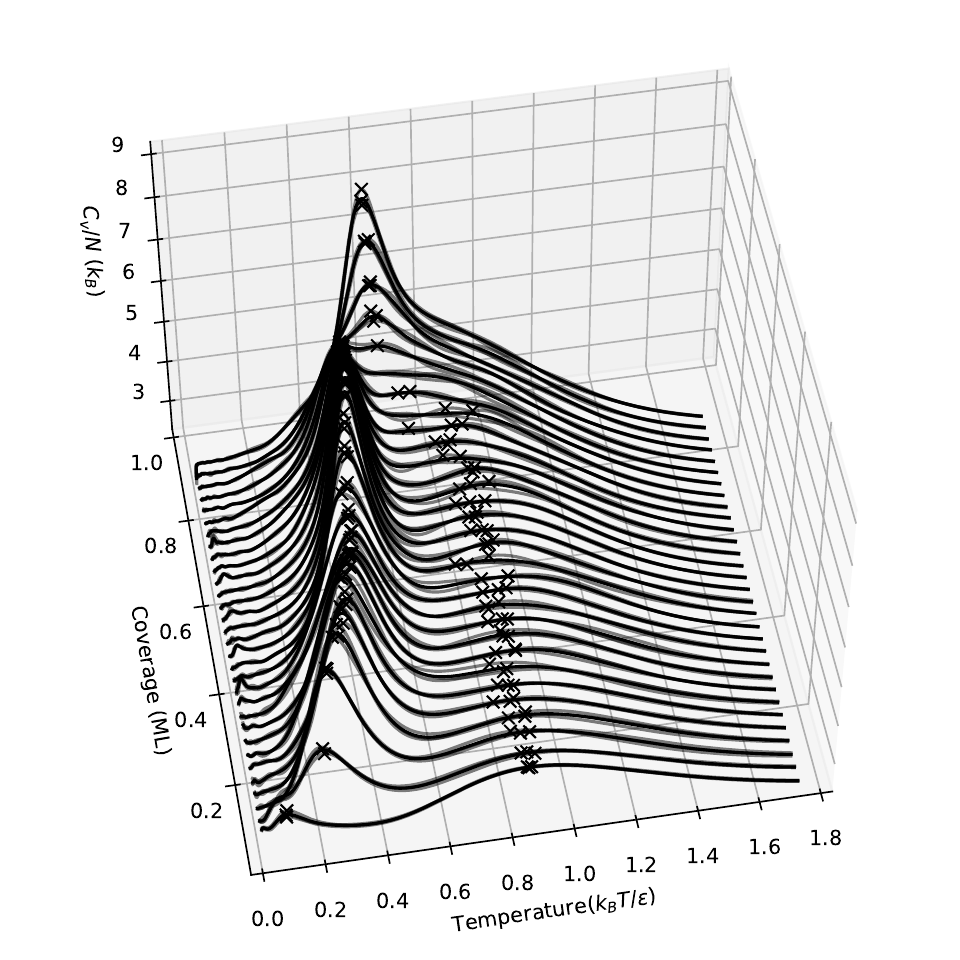}
        \label{fig:surf111_6x6_cv}
    \end{subfigure}
    \begin{subfigure}[t]{0.43\textwidth}
        \caption{}
        \includegraphics[width=\textwidth]{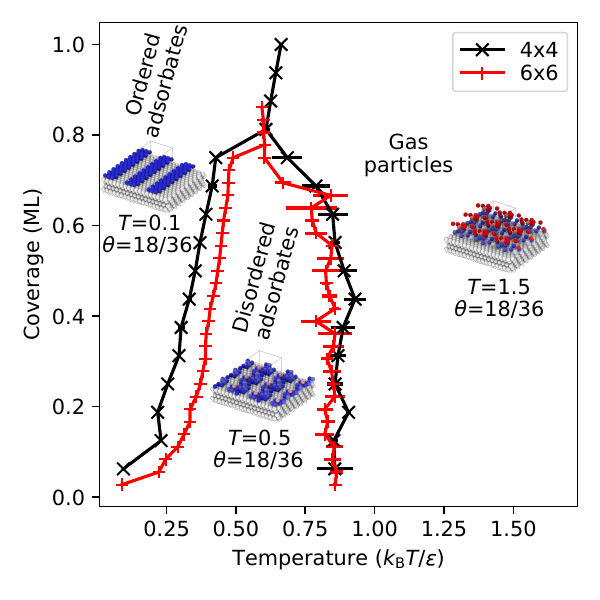}
        \label{fig:surf111_6x6_phase_diagram}
    \end{subfigure}
    \caption{
Coverage-temperature properties for the flat Lennard-Jones(111) surface are presented as follows: (a) heat capacity per free particle on a $2\times2$ surface with fractional coverages ranging from $\theta=1/4$ ML to three MLs; (b) phase diagram for the same $2\times2$ surface; (c) heat capacity per free particle on an expanded $6\times6$ surface with fractional coverages ranging from $\theta=1/36$ ML to $\theta=31/36$ ML; and (d) the phase diagram for this $6\times6$ surface (in red), with insets showing the maximum-probability structures of each phase at selected temperatures for $\theta=18/36$~ML, juxtaposed with the phase diagram for the $4\times4$ surface as shown in Figure~\ref{fig:surf111_phase_diagram} (in black). The $6\times6$ unit cell is repeated three times in $x$- and $y$-directions for better visibility of the surface structures. The lines between the crosses are only guides for the eye.
    }
    \label{fig:111-2x2-cv}
\end{figure*}

We first examine how the calculated thermodynamic properties change on a smaller $2\times2$ per monolayer (ML) unit cell, where the finite-size effects should be more pronounced. Additionally, a smaller unit cell allows us to increase the number of free particles in the system to form more than one ML. The results are presented in Figures~\ref{fig:surf111_2x2_cv} and \ref{fig:surf111_2x2_phase_diagram}.

One can see that the finite-size effect is significant. The lower-temperature, entropy-induced peaks in the heat capacity curves are not found (besides the one- and two-free-particle cases, $\theta=1/4$ and $2/4$ ML), which is intuitive, as the number of configurations on a $2\times2$ surface is significantly lower than the one with $4\times4$ surface particles. Nevertheless, the smaller system allows us to examine the phase transitions with surface coverages beyond one ML, where the transition temperatures exhibit weak coverage dependency.

To verify the accuracy of our $4\times4$ surface results and to ensure minimal finite-size effects, we performed nested sampling (NS) calculations using a larger $6\times6$ particles-per-ML surface unit cell. These calculations are computationally demanding due to the inclusion of more free particles for equivalent fractional coverages and a significantly larger sampling volume. For efficient phase-space sampling, we utilized 256 walkers per free particle. Our calculations converged to a fractional coverage of $\theta=31/36$ ML, necessitating roughly three million NS iterations to reach the lowest potential energy state.

The heat capacity curves derived from these calculations can be found in Figure~\ref{fig:surf111_6x6_cv}, with the corresponding phase diagram in Figure~\ref{fig:surf111_6x6_phase_diagram}. The results for the $4\times4$ surface align closely with those for the $6\times6$ surface, both qualitatively and quantitatively. They produce similar phase diagrams with consistent transition temperatures and a triple point. Additionally, the surface structures observed on the $4\times4$ and $6\times6$ surfaces at equivalent fractional coverages verify the adequacy of the $4\times4$ surface size. In conclusion, the more computationally modest $4\times4$ surface effectively captures the relevant physics. This computational efficiency enables us to explore a broader spectrum of surfaces beyond Lennard-Jones(111) [LJ(111)], making the $4\times4$ surface our system size of choice.

\subsection{Additional details on surface system setup} \label{sec:surface-setup-details}

In this section, we compare the computational outputs from simulations with different system setups, illustrating the importance of having the appropriate configuration for the sampled system.

\begin{figure*}[htb]
    \centering
    \includegraphics[width=\textwidth]{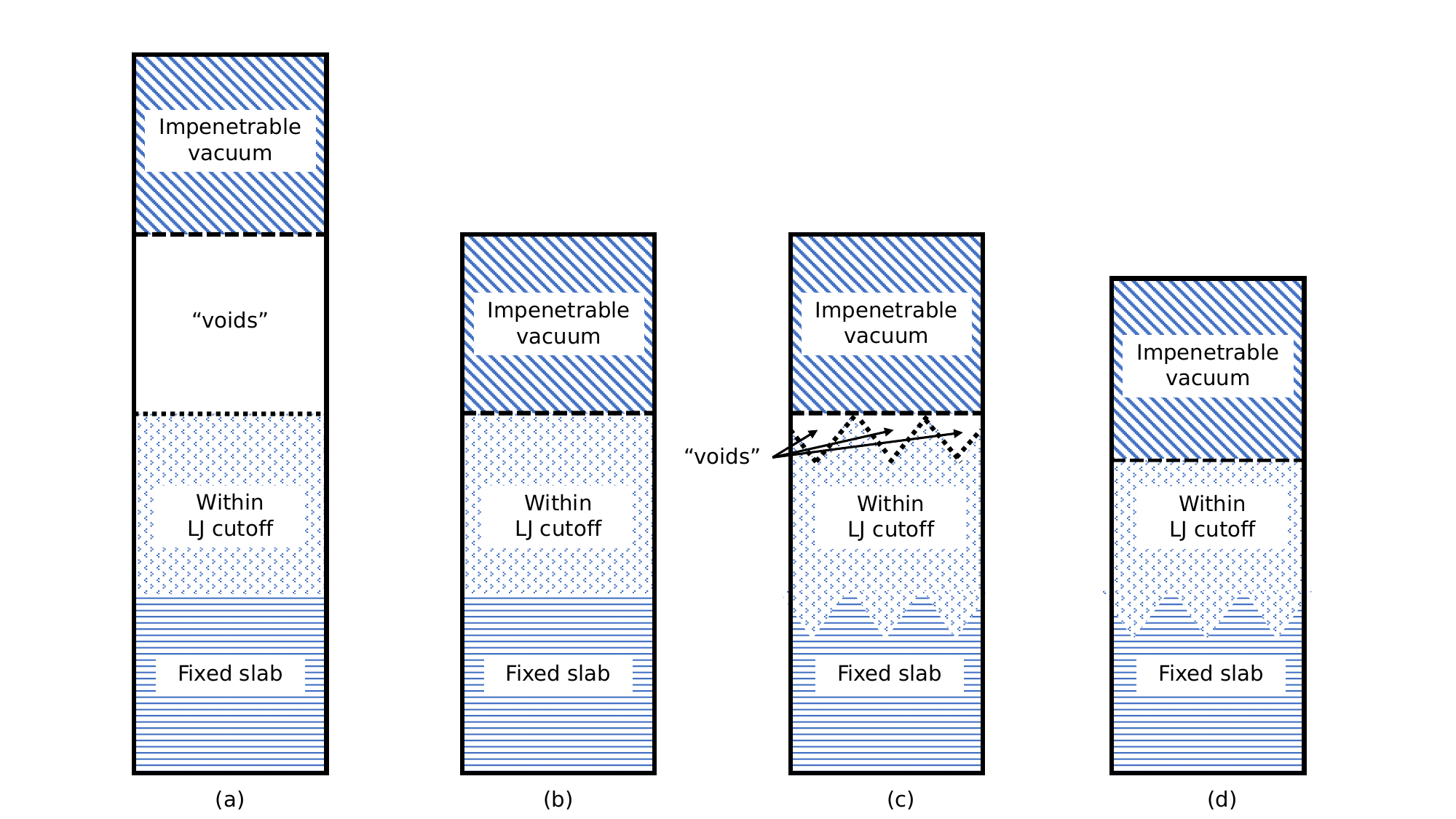}
    \caption{
Different ways to construct the fixed surface-free particles model: The key is to ensure that all free particles can interact with the surface. In setup (a), there is a large volume outside the LJ cutoff, labeled as ``voids'', where the free particles cannot interact with the fixed surface. Such a setup is considered problematic in this work, as we may accidentally sample the clustering of particles in the ``voids''. Setup (b) is appropriate for flat surfaces. The ``voids'' are removed by reducing the overall height of the cell. Such a setup is used for LJ(111) and LJ(100) surfaces. However, even with reduced cell height, significant ``voids'' can still appear, as shown in (c), with stepped surfaces such as LJ(311) and LJ(110). We further reduced the cell's height to correct it, as shown in (d).
    }
    \label{fig: LJ-scheme}
\end{figure*}

To adequately capture the interplay between free particles and the host surface, we must ensure that the LJ interactions from the surface cover the entire volume of the available space. Otherwise, particles may move into regions with no interactions during sampling, leading to either no potential energy contribution or inaccurate contributions from particle-particle interactions rather than particle-surface interactions. Figure~\ref{fig: LJ-scheme} depicts several scenarios of system setups that lead to different final results.

\begin{figure}[tb]
    \centering
    \includegraphics[width=0.5\textwidth]{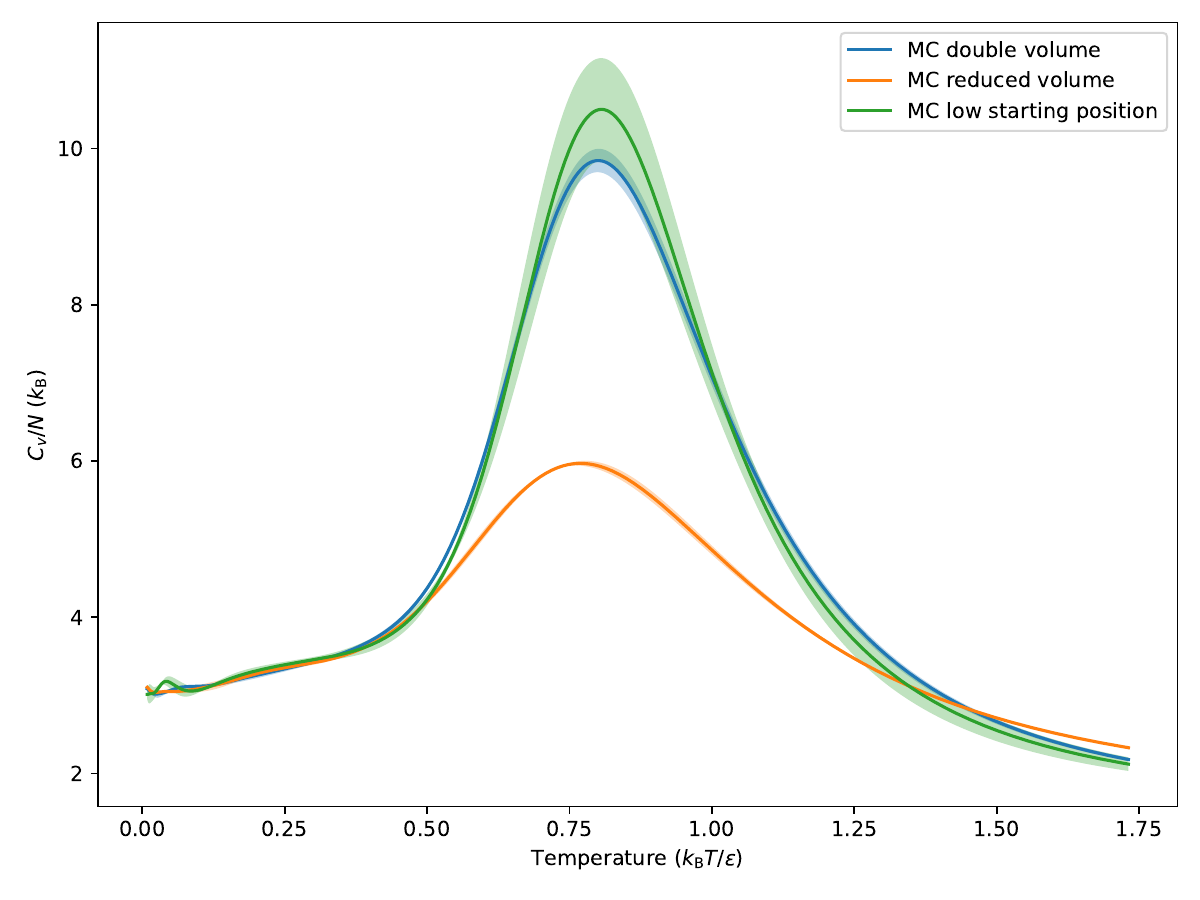}
    \caption{
Heat capacity of a $2\times2$ LJ(111) surface with three free particles. The ``double volume'' (blue) curves were produced from NS using the setup shown in Figure~\ref{fig: LJ-scheme}(a). The ``reduced volume'' curves were produced using the setup shown in Figure~\ref{fig: LJ-scheme}(b). The ``low starting positions'' (green) curves were generated by utilizing the setup in Figure~\ref{fig: LJ-scheme}(a), but all initial walkers were placed within the LJ cutoff range. Although the overall volume remains the same in this setup, all free particles can interact initially with the surface. Only the ``reduced volume'' (orange) curves are considered for this study.
    }
    \label{fig: LJ_3p_setup}
\end{figure}

We have conducted tests on how the presence of ``voids'' and the placement of initial walkers influence the NS process and the calculated heat capacities. Three tests were carried out on a $2\times2$ LJ(111) surface with three free particles. The results are summarized in Figure~\ref{fig: LJ_3p_setup}. The first set of tests used the ``double volume'' setup, as shown in Figure~\ref{fig: LJ-scheme}(a). The second set of tests utilized the ``reduced volume'' setup, as displayed in Figure~\ref{fig: LJ-scheme}(b). The third set of tests used the setup shown in Figure~\ref{fig: LJ-scheme}(a) but placed all initial walkers within the LJ cutoff range (labeled as ``low starting positions'' in Figure~\ref{fig: LJ_3p_setup}), where the overall volume is the same and all free particles can initially interact with the surface. However, this setup does not have a uniform distribution of initial walkers in the configuration space. We observed that the overall volume determines the height of the peak in the $C_V$ curves and has minimal influences on the transition temperature. This volume dependence is unsurprising as the difference in volume leads to changes in other conditions, \textit{e.g.} pressure, thus altering the $C_V$. Interestingly, the ``low starting positions'' setup recovers the same $C_V$, albeit with significant variance, even though a non-uniform distribution of initial walkers is used. We consider the setup used to produce the orange curves in Figure 1 appropriate for this work.

\begin{figure}[tb]
    \centering
    \includegraphics[width=0.5\textwidth]{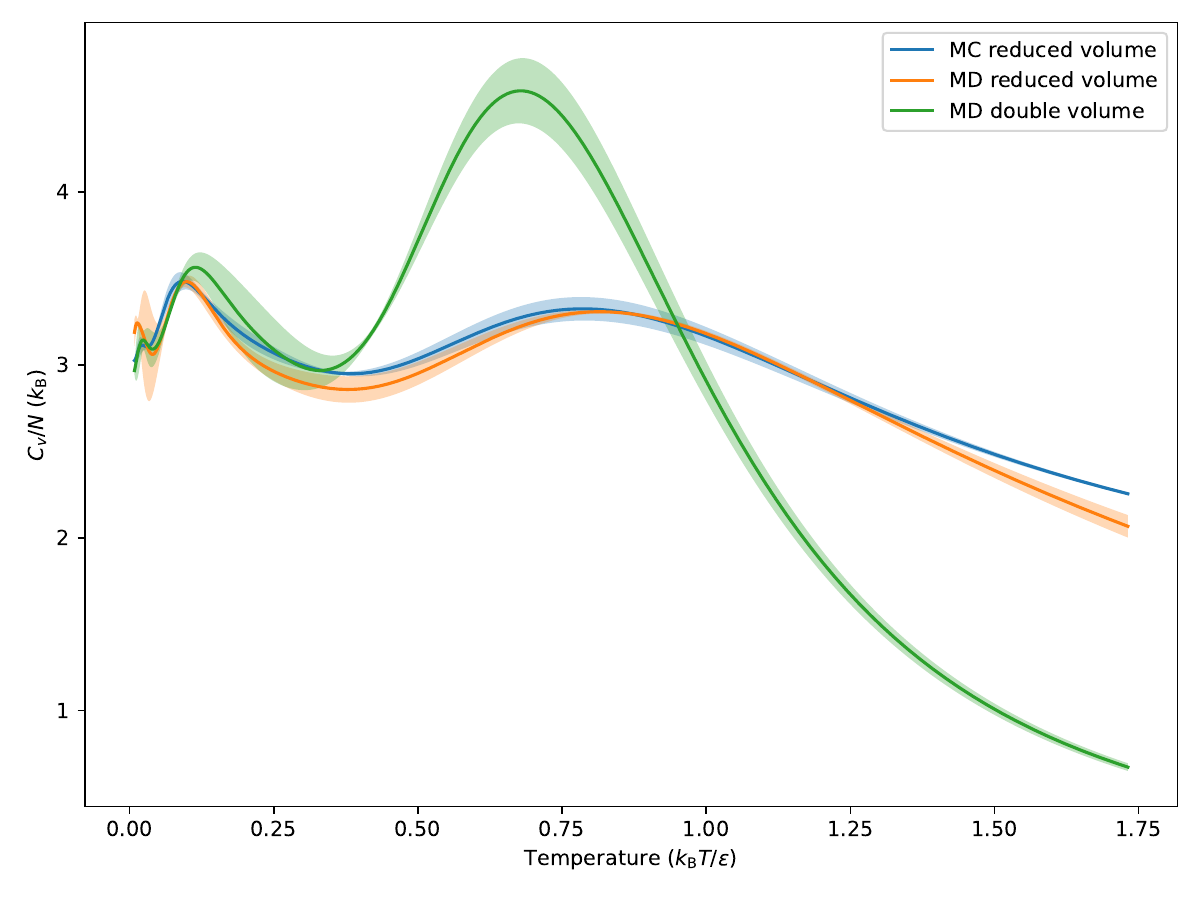}
    \caption{
Heat capacity of a $2\times2$ LJ(111) surface with a single free particle. ``MD'' in the legend indicates that the free particle is propagated using molecular dynamics (MD) in the Large-scale Atomic/Molecular Massively Parallel Simulator (LAMMPS). \cite{thompson_lammps_2022} The NS runs with the ``double volume'' setup, as shown in Figure~\ref{fig: LJ-scheme}(a), get stuck whenever the single particle is in a ``void.'' The ``reduced volume'' curves, shown in blue, are from the setup as shown in Figure~\ref{fig: LJ-scheme}(b). These are the correct results, as verified by the MD (represented in orange) results, using LAMMPS with an identical system setup.
    }
    \label{fig: LJ_1p_setup}
\end{figure}

Furthermore, we encountered a technical challenge when running NS for any surfaces with a single free particle using the ``double volume'' setup, as shown in Figure~\ref{fig: LJ-scheme}(a). The sampling can stall when the free particle is placed in the ``void'' region, either by a Monte Carlo (MC) move or generated as an initial walker. As the ``void'' region is rather large and there are no other particles with which to interact, generating a new configuration with lower potential energy using random MC moves is challenging due to the small step size relative to the cell's dimensions. Currently, \verb|pymatnest| updates the MC step size to maintain an acceptance ratio of 0.25-0.75. However, when particles enter these voids during MC steps, the potential energy does not change. This condition decreases the acceptance ratio because we only accept the MC step if the potential energy remains below $E_i^{\max}$. As a result, adjusting the MC step size to infinitesimal values becomes necessary to restore the acceptance ratio to 0.25 and 0.75. NS with molecular dynamics can proceed with the ``double volume'' setup (green curves in Figure~\ref{fig: LJ_1p_setup}) via $NVE$ propagation. As discussed above, the phase transitions, especially the higher-temperature condensation processes, change significantly with additional volume. To produce consistent $C_V$ with only MC propagation, we prefer eliminating any additional vacuum beyond the LJ cutoff range.

\subsection{Surface structures}

In Table~\ref{tab: SI_facets_params}, we report the parameters of the four facets of the LJ solid. The cell dimensions refer to the dimensions of the simulation cell used in the NS calculations. The vacuum thickness is the distance between the topmost fixed layers and the reflective wall (at $4\sigma$ below the top of the simulation cell), and the thickness is less than or equal to the LJ potential cutoff. The slab thickness is the distance between the topmost fixed layers and the bottom of the cell. For each facet, there are different numbers of layers depending on the interlayer spacing. The trough spacing is the distance between the centers of two neighboring troughs, applicable for the (110) and (311) facets.

\begin{table*}[htb]
    \centering
    \begin{tabular}{ccccccc}
        \hline
        Facet & Cell dimensions & Vacuum thickness & Slab thickness & Number of layers & Interlayer spacing & Trough spacing \\
        \hline
        (111) & $4.49 \times 3.89 \times 11.66$ & 4.00 & $3.66$ & $5$ & $0.92$ & $-$ \\
        (110) & $6.34 \times 4.49 \times 11.37$ & 3.44 & $3.93$ & $8$ & $0.56$ & $1.60$ \\
        (100) & $4.49 \times 4.49 \times 11.97$ & 4.00 & $3.97$ & $6$ & $0.79$ & $-$ \\
        (311) & $4.49 \times 7.44 \times 11.35$ & 3.52 & $3.83$ & $9$ & $0.48$ & $1.86$
    \end{tabular}
    \caption{
The parameters of the four facets of the LJ solid are all in units of $\sigma$. They include cell dimensions, represented by $x \times y \times z$; the vacuum thickness, which is the distance between the topmost fixed layers and the reflective wall; the slab thickness, defined as the distance between the topmost fixed layers and the bottom of the cell; the number of layers within the fixed slab; the interlayer spacing, which is the distance between two layers in the fixed slab; and the trough spacing, the distance between the centers of two neighboring troughs.
    }
    \label{tab: SI_facets_params}
\end{table*}

\subsection{Phase diagrams for flat LJ(100) and stepped LJ(311) surfaces}

\subsubsection{Flat LJ(100) surface}

\begin{figure*}[htbp]
    \centering
    \begin{subfigure}[t]{0.45\textwidth}
        \caption{}
        \includegraphics[width=\textwidth]{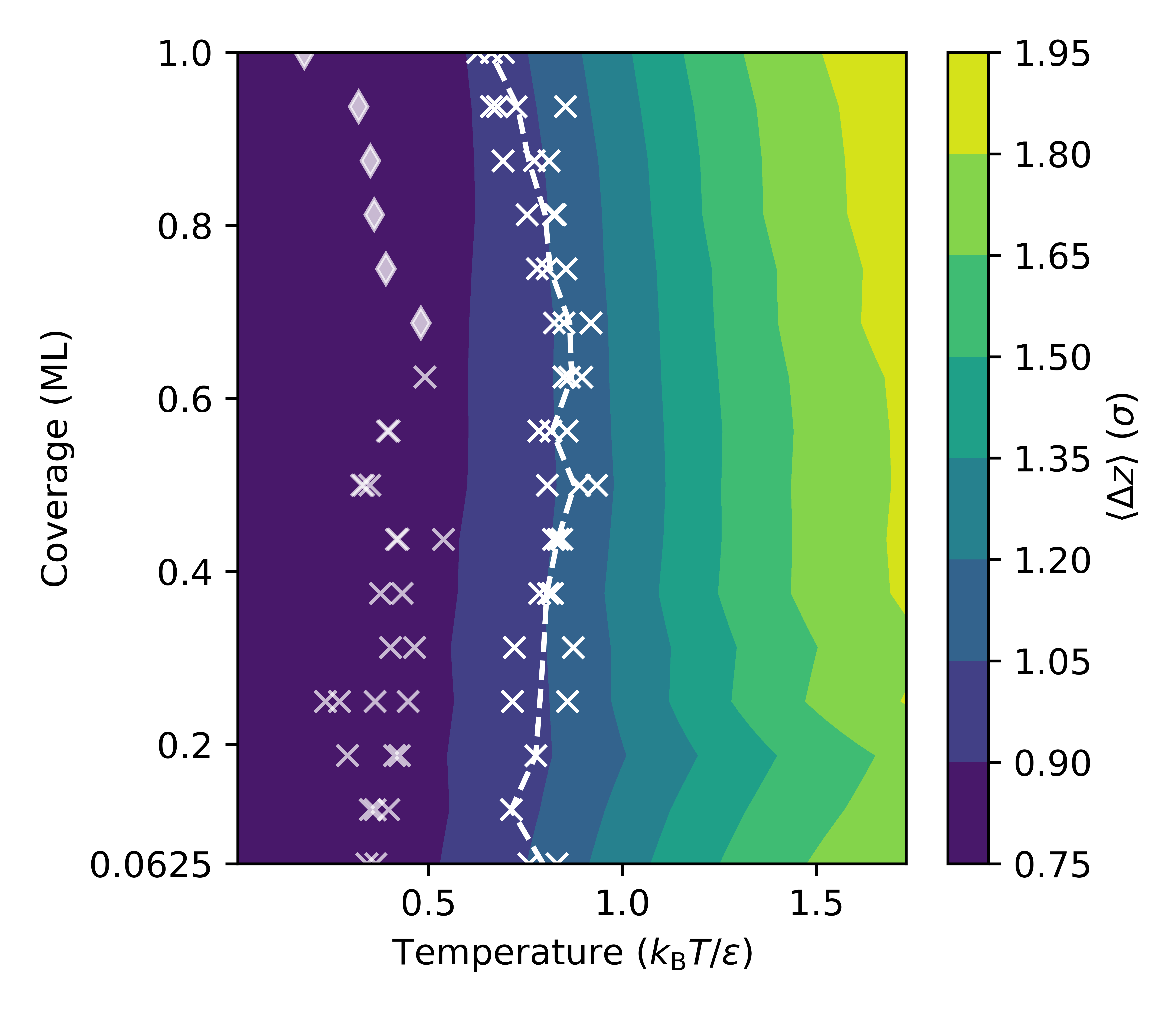}
        \label{fig:surf100_Z}
    \end{subfigure}
    \begin{subfigure}[t]{0.45\textwidth}
        \caption{}
        \includegraphics[width=\textwidth]{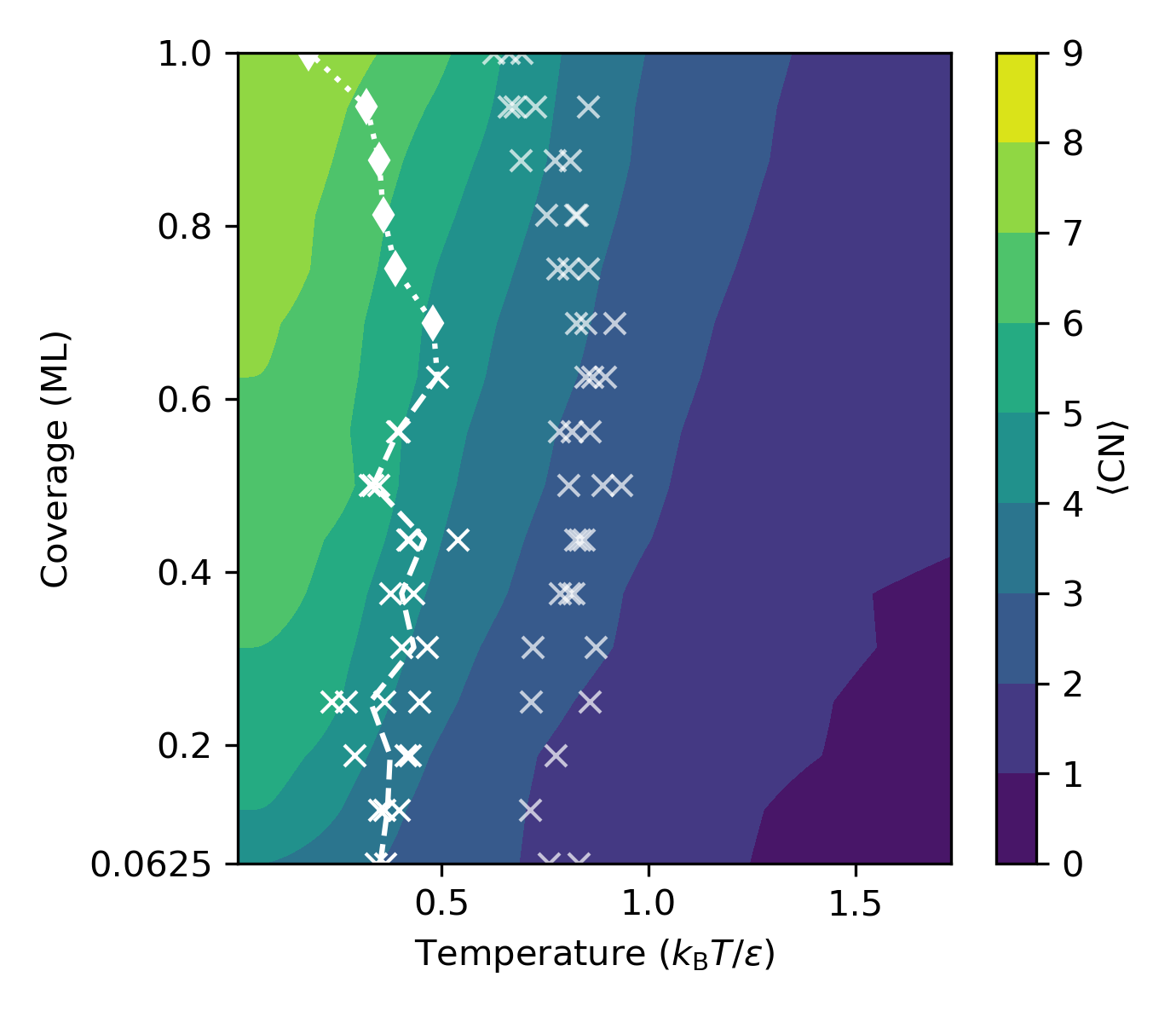}
        \label{fig:surf100_CN}
    \end{subfigure}
    \caption{
Calculated coverage-temperature properties of the flat LJ(100) surface with fractional coverages from $\theta=1/16$ ML up to one ML. (a) The average $z$-coordinates of the free particles are shown relative to the topmost layer in the fixed slab. Note that the bulk (100) interlayer spacing is $0.79\sigma$. (b) The free particles' average coordination numbers, including particle-particle and particle-surface bonding, are indicated. The lines between the crosses are just guides for the eye. The diamond markers ($\blacklozenge$) point to the disappearing peaks not found by the automated procedure.
    }
    \label{fig:100-4x4-cv}
\end{figure*}

As the LJ(100) and LJ(111) surfaces share similarities in terms of their surface features and phase behaviors, we only present the LJ(111) results in the main text. Most of the phase behaviors of the LJ(100) surface can be understood following the discussions of the LJ(111) surface. Here, we only highlight specific differences between the two surfaces. Compared to LJ(111), LJ(100) is also considered a flat surface but with reduced surface symmetry (four-fold \textit{vs.}~six-fold). This reduction leads to broader and lower low-temperature peaks in the $C_V$ curves (Figure~\ref{fig:cv100}) for coverages $\theta\leq10/16$ ML. For $\theta>10/16$ ML, the low-$T$ peaks disappear. We manually determine a ``shoulder'' peak for each $C_V$ curve with  $\theta>10/16$ ML and mark them with $\blacklozenge$ in Figure~\ref{fig:phase100}. One can see that heat capacity peaks for the entropy-driven ordering phase transition are being diminished as coverage increases beyond the half-ML ($\theta=8/16$ ML), where the coordination number is being maximized. Note that the maximum coordination number for surface particles in a (100) ML is eight, with four from the slab and four from neighboring particles. Significantly, the triple point is not present in the LJ(100) phase diagram (Figure~\ref{fig:phase100}), which is present in the LJ(111) phase diagram. This absence indicates that, on LJ(111), for high surface coverages, the entropic and enthalpic effects compete, whereas, on LJ(100), the entropy-driven (lower temperature) ordering phase transition disappears while the enthalpy-driven (higher temperature) condensation phase transition remains unaffected.

\subsubsection{Stepped LJ(311) surface}

\begin{figure*}[htbp]
    \centering
    \begin{subfigure}[t]{0.45\textwidth}
        \caption{}
        \includegraphics[width=\textwidth]{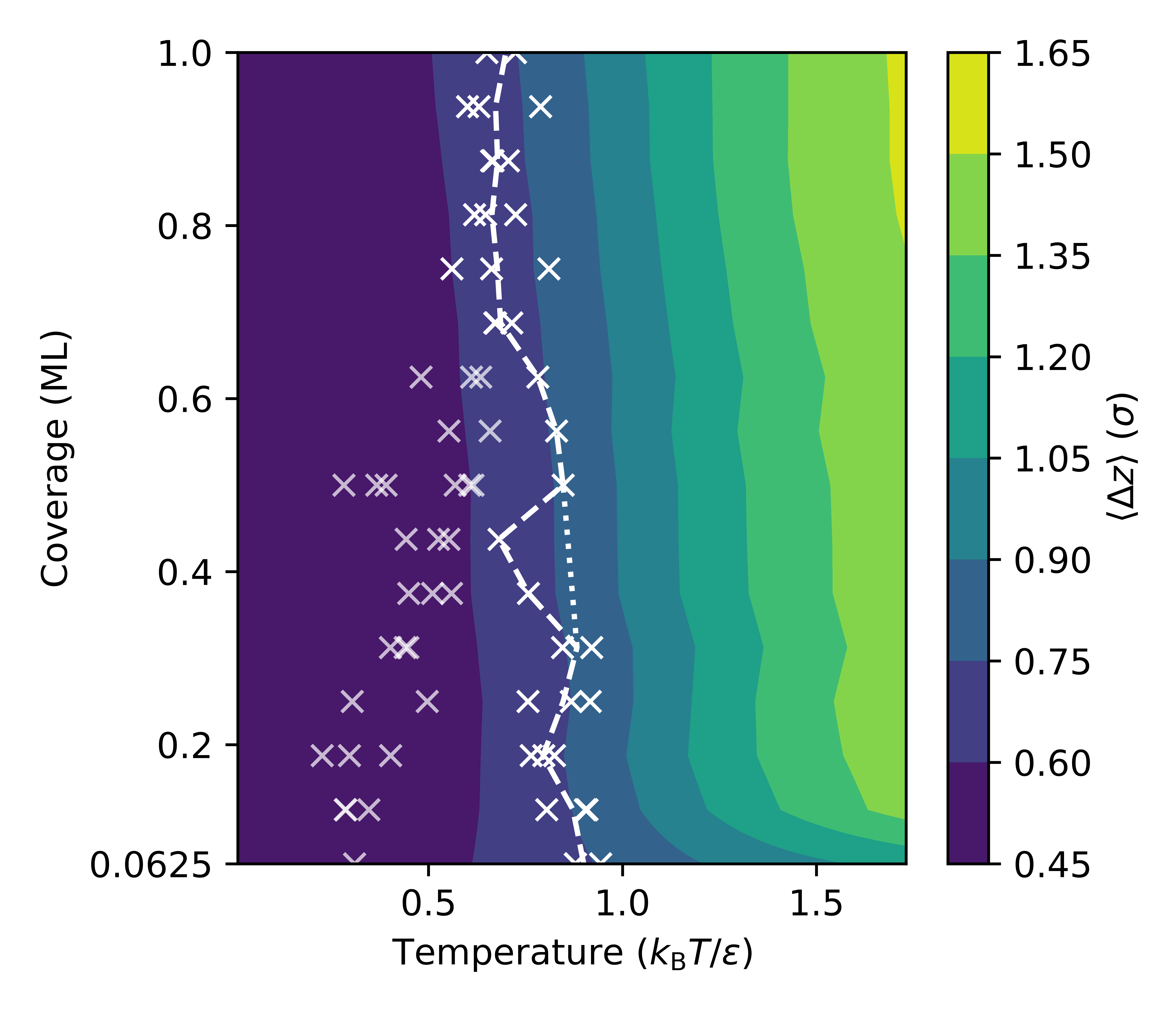}
        \label{fig:surf311_Z}
    \end{subfigure}
    \begin{subfigure}[t]{0.45\textwidth}
        \caption{}
        \includegraphics[width=\textwidth]{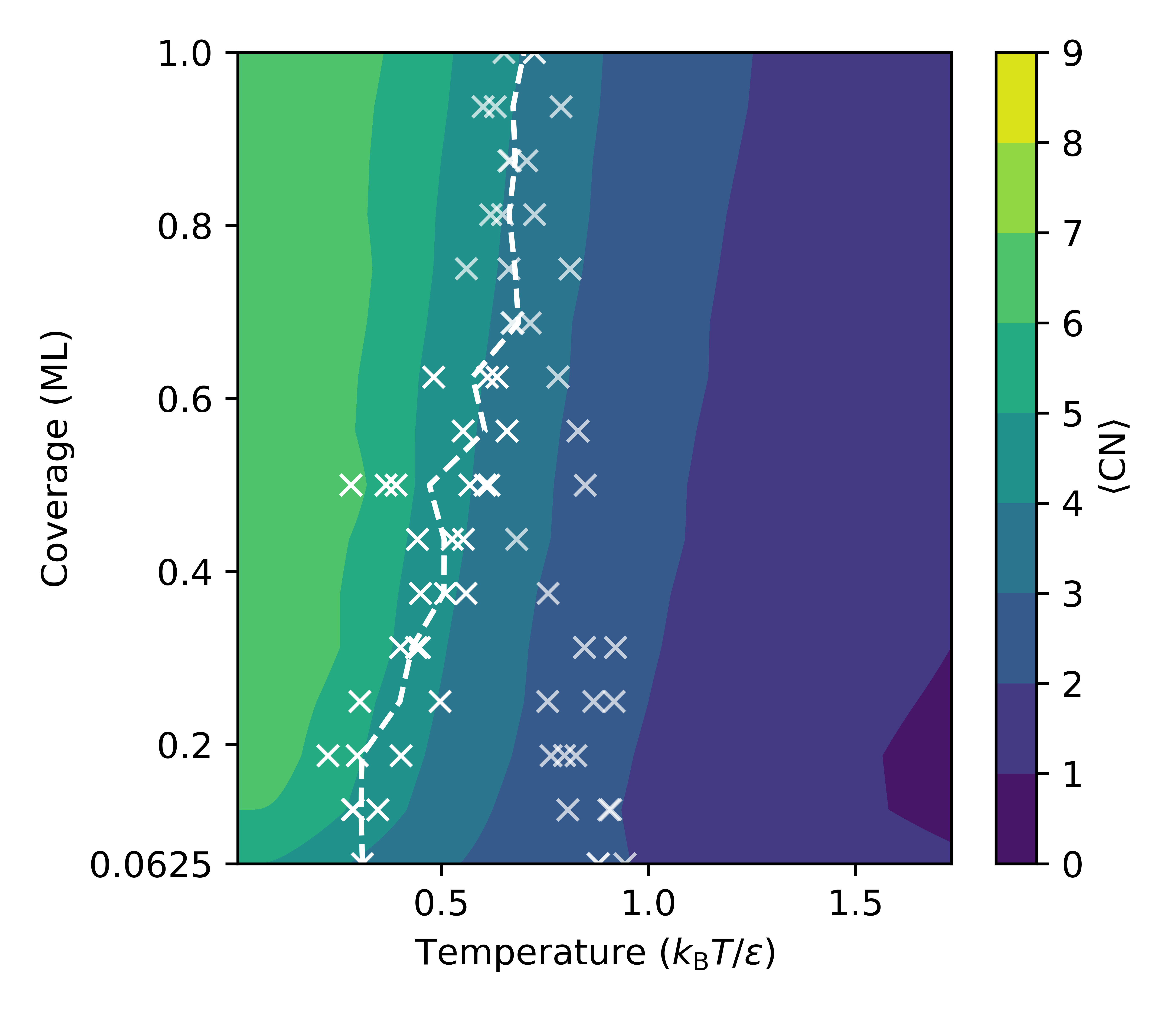}
        \label{fig:surf311_CN}
    \end{subfigure}
    \caption{
Calculated coverage-temperature properties of the stepped LJ(311) surface with fractional coverages from $\theta=1/16$ ML up to one ML. (a) The average $z$-coordinates of the free particles are shown relative to the topmost layer in the fixed slab. Note that the bulk (311) interlayer spacing is $0.48\sigma$. (b) The free particles' average coordination numbers, including particle-particle and particle-surface bonding, are indicated. The lines between the crosses are just guides for the eye.
    }
    \label{fig:311-4x4-cv}
\end{figure*}

The stepped surface with a higher Miller index has surface features such as ``troughs'' along one of the lateral directions, offering several different binding sites compared to the flat LJ(111) facet. The potential energy landscape of LJ(311) is more complex than LJ(111), and its adsorbate phase diagram is more difficult to determine. We compute the $C_V \left( T \right)$ for the LJ(311) surface with the same coverages as those for the flat LJ(111) surface. The $C_V \left( T \right)$ curves are shown in Figure~\ref{fig:cv311}.

Interpreting the $C_V \left( T \right)$ curves for the LJ(311) surface is more challenging than those for the LJ(111) surface. First, the LJ(311) $C_V \left( T \right)$ curves from the NS runs show larger deviations near the peaks than those for LJ(111). Moreover, the peaks are generally broader and significantly overlap, making them less precise. For example, at $\theta=6/16$ and $7/16$, the $C_V$ curves have a plateau-like feature from $T=0.5-0.8~k_{\mathrm{B}} T / \epsilon$. Despite these difficulties, we utilized the same automated procedure to find the peaks. We noted that for each of the lower coverages ($\theta\leq10$), there are two separate peaks: one at a higher temperature between $0.7$ and $0.9~k_{\mathrm{B}} T / \epsilon$ (see $\times$s), and another peak at a lower temperature between $0.2$ and $0.6~k_{\mathrm{B}} T / \epsilon$. Compared to the $C_V$ curves of LJ(111), LJ(311)'s low-coverage, high-temperature peaks are more dominant.

For higher coverages $(\theta\geq11/16)$, the peaks on the $C_V$ curves have merged into one, similar to the behavior observed on LJ(111) at higher coverages. The merged peaks are typically located at a temperature between $0.6$ and $0.8~k_{\mathrm{B}} T / \epsilon$ and are often broader and lower than those for LJ(111).

We further project the temperatures of all peaks onto the coverage-temperature plot to construct the phase diagram for LJ(311), as shown in Figure~\ref{fig:phase311}. Due to the more significant deviations of the peak temperatures from the $C_V$ curves, the phase boundaries are less clearly defined than those for LJ(111). That said, the overall features of the LJ(311) phase diagram are still qualitatively similar to LJ(111): two phase transitions for coverages $\theta\leq10$, a triple point located at $T \approx 0.7~k_{\mathrm{B}} T / \epsilon$ and $\theta=11/16$, and a single phase transition for coverages $\theta\geq12/16$. Note that, due to the flatness of the $C_V$ curves at $\theta=6/16$ and $7/16$, the peaks found by the \verb|scipy.signal.find_peaks()| function may not be reliable. Therefore, we join high-temperature peaks at $\theta=5/16$ and $8/16$, marked by the dotted line in Figure~\ref{fig:phase311}, to show a more likely phase boundary.

In order to further understand the phase transitions on the LJ(311) surface, we also computed surface order parameters, including $\langle \Delta z \rangle$ (Figure~\ref{fig:surf311_Z}) and $\langle \mathrm{CN} \rangle$ (Figure~\ref{fig:surf311_CN}). We can utilize the results from LJ(111) and LJ(110) to infer the phase transitions on the LJ(311) surface. LJ(311) can be understood as an intermediate surface between the completely flat LJ(111) and the stepped LJ(110) surfaces. The troughs on LJ(311) are shallower than LJ(110), and the inter-trough spacing is larger (see Table~\ref{tab: SI_facets_params}), hence it is more ``flat.''

\subsection{Stepped LJ(110) surface} \label{sec: 110}

\begin{figure*}[htb]
    \centering
    \begin{subfigure}[t]{0.45\textwidth}
        \caption{}
        \includegraphics[width=\textwidth]{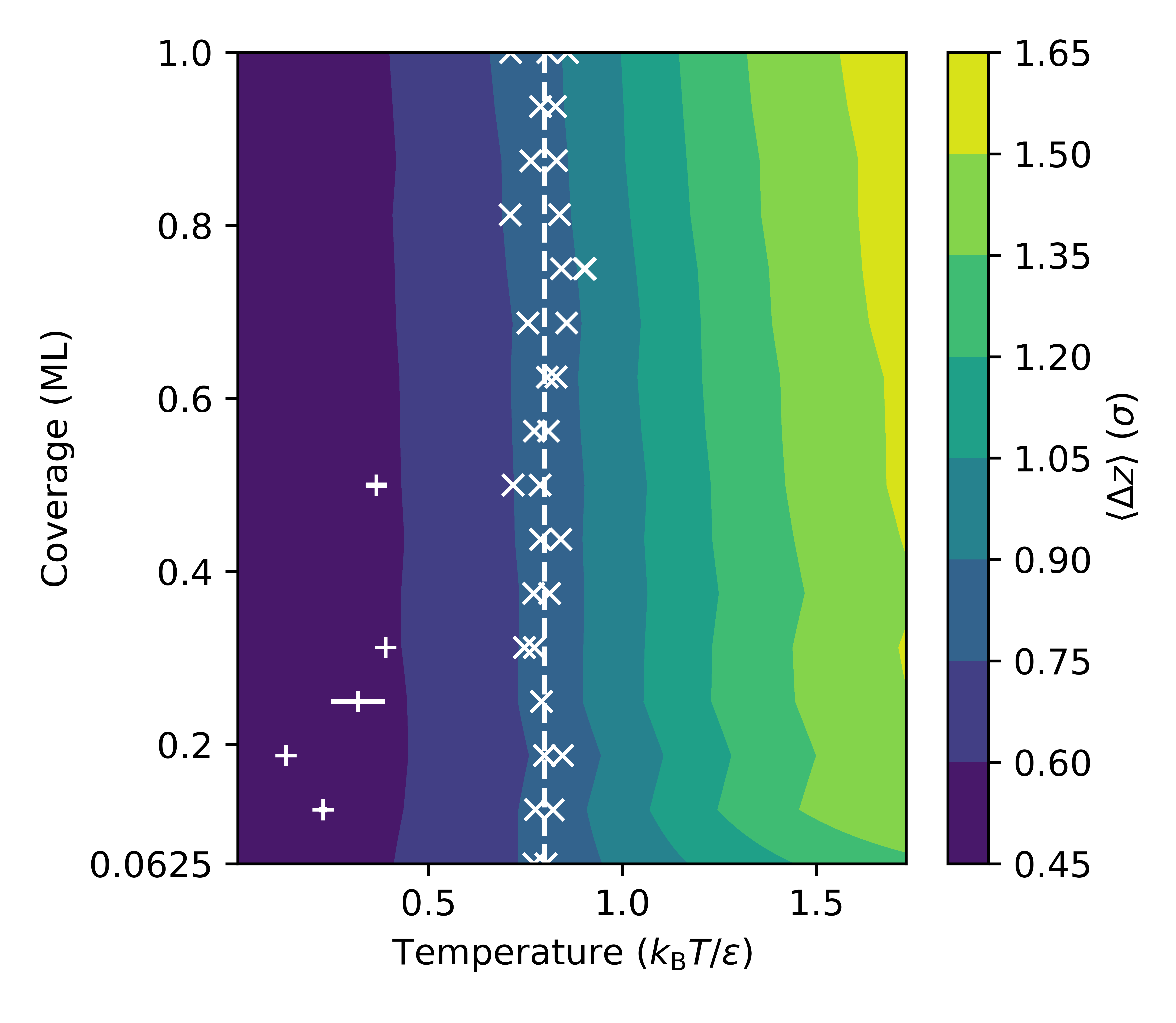}
        \label{fig:surf110_Z}
    \end{subfigure}
    \begin{subfigure}[t]{0.45\textwidth}
        \caption{}
        \includegraphics[width=\textwidth]{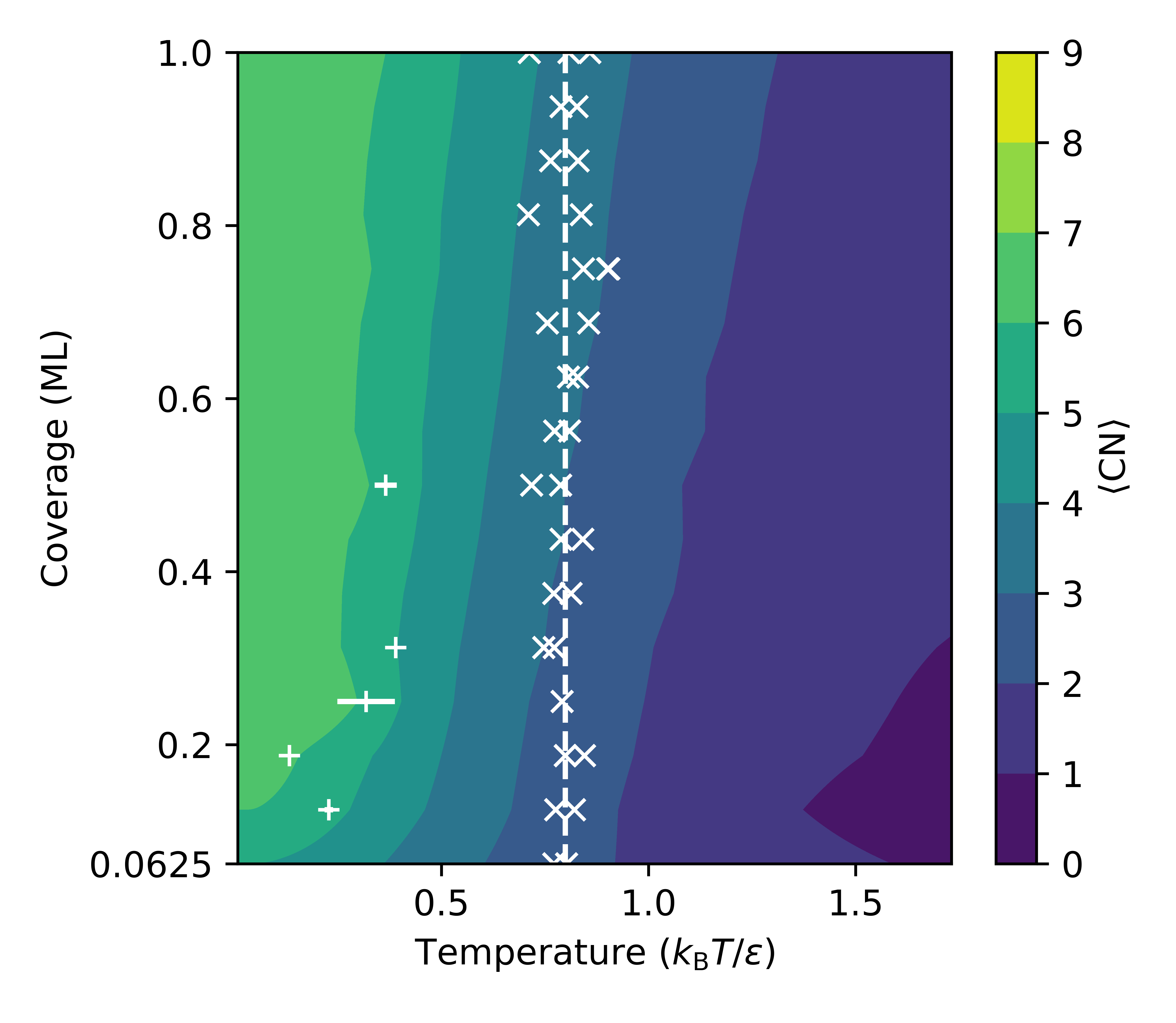}
        \label{fig:surf110_CN}
    \end{subfigure}
    \caption{
Calculated coverage-temperature properties of the stepped LJ(110) surface with fractional coverages from $\theta=1/16$ ML up to one ML. (a) The average $z$-coordinates of the free particles are shown relative to the topmost layer in the fixed slab. Note that the bulk (110) interlayer spacing is $0.56\sigma$. (b) The free particles' average coordination numbers, including particle-particle and particle-surface bonding, are indicated. The dashed line at $\overline{T}_\theta=0.80~k_{\mathrm{B}} T / \epsilon$ shows the coverage-averaged phase transition temperature.
    }
    \label{fig:110-4x4-cv}
\end{figure*}

We compute $C_V \left( T^* \right)$ for the stepped LJ(110) surface with the same coverages as those for the flat LJ(111) surface. We refer to the LJ(110) surface as stepped because it has ``troughs'' along one of the lateral directions. The $C_V$ curves in Figure~\ref{fig:cv110} show a dominant and almost coverage-independent peak (using the same automated procedure as that used in Section~\ref{sec: 111}) at temperatures between $0.7~k_{\mathrm{B}} T / \epsilon$ and $0.9~k_{\mathrm{B}} T / \epsilon$ (see $\times$s). Note that a side peak can be observed at certain lower coverages (\textit{e.g.}, $\theta=2/16$ ML, $\theta=4/16$ ML, and $\theta=8/16$ ML), but no clear trend emerges. For the coverages without a visible side peak, such peaks may be masked by the dominant peaks. We do not yet have a reliable tool to resolve them, and we wish to develop a technique for masked peak detection in the future. The dominant and almost coverage-independent peak corresponds to an adsorbate phase transition, with a coverage-averaged phase transition temperature, $\overline{T}_\theta$, of $0.80(3)~k_{\mathrm{B}} T / \epsilon$, as indicated by the dashed line in Figure~\ref{fig:phase110}.

We calculated $\langle \Delta z \rangle$ to characterize the adsorbate phases and their transitions. Figure~\ref{fig:surf110_Z} shows that, at the highest temperature considered, \textit{i.e.}, $T \approx 1.73~k_{\mathrm{B}} T / \epsilon$, $\langle \Delta z \rangle = 1.65 \sigma$ is nearly equal to the vertical center of \textbf{Region 2} in the simulation cell, \textit{i.e.}, $1.72 \sigma$ (See Table~\ref{tab: SI_facets_params} in the ESI). At $\overline{T}_\theta$, $0.75\sigma < \langle \Delta z \rangle < 0.90\sigma$. Note that in a perfect LJ(110) bulk,  the $\Delta z_{\mathrm{bulk}} = 0.84\sigma$ (the average of the trough site at $0.56\sigma$ and the atop site at $1.12\sigma$). This correspondence between $\langle \Delta z \rangle$ and $\Delta z_{\mathrm{bulk}}$ suggests that the free particles ``condense'' on the stepped LJ(110) about one interlayer spacing above the surface, which we also observed on the LJ(111) in the previous section.

However, the nature of surface condensation on LJ(110) differs from that on LJ(111), as seen from the widths of their $C_V$ peaks. To further characterize surface condensation, we calculate $\langle \mathrm{CN} \rangle$ (see Figure~\ref{fig:surf110_CN}), which is approximately three at $\overline{T}_\theta$. The average is derived from two scenarios: two free-fixed bonds when an adsorbed particle touches only one side of the trough and four when the particle touches both sides but has not fully settled into position. The coordination is at least fivefold when the particles are completely situated within the trough sites because they bind the fixed surface particles from two distinct layers, forming a square pyramidal structure. At near-zero temperature, $\langle \mathrm{CN} \rangle$ is five for $\theta=1/16$ ML (\textit{i.e.}, five free-fixed bonds), six for $\theta=2/16$ ML (\textit{i.e.}, one free-free and five free-fixed bonds), and seven for $\theta\geq3/16$ ML (\textit{i.e.}, two free-free and five free-fixed bonds). Figure~\ref{fig:surf110_CN} also indicates that the free particles prefer to occupy the same trough if it has at least one unoccupied site because the maximum $\langle \mathrm{CN} \rangle$ for $\theta\geq3/16$ ML is seven, which is not possible if the free particles occupy different troughs.

\twocolumn[
  \begin{@twocolumnfalse}

\subsection{Maximum-probability structures}\label{sec: SI-structures}

We include top and side views of all four different facets on $4\times4$ surfaces at selected temperatures at all coverages as separate files:
  
\begin{itemize}
    \item LJ(111) surface: \texttt{111-4x4-grid-top.png} and \texttt{111-4x4-grid-side.png}
    \item LJ(311) surface: \texttt{311-4x4-grid-top.png} and \texttt{311-4x4-grid-side.png}
    \item LJ(100) surface: \texttt{100-4x4-grid-top.png} and \texttt{100-4x4-grid-side.png}
    \item LJ(110) surface: \texttt{110-4x4-grid-top.png} and \texttt{110-4x4-grid-side.png}
\end{itemize}

We also include the top and side views of the (111) facet on the $6\times6$ at selected temperatures at coverages up to $\theta=31/36$ ML as separate files:

\begin{itemize}
    \item LJ(111) surface: \texttt{111-6x6-selected-grid-top.png} and \texttt{111-6x6-selected-grid-side.png}
\end{itemize}

\end{@twocolumnfalse} \vspace{0.6cm}
]
\end{document}